\documentclass[10pt,twoside,preprint2]{aastex}

\usepackage{natbib}
\usepackage{graphicx} 
\usepackage{subfigure}

\usepackage[scaled=1.00]{helvet}

\usepackage[T1]{fontenc}
\usepackage{textcomp}

\usepackage{ragged2e} 

\usepackage{sectsty}
\sectionfont{\normalsize}
\subsectionfont{\normalsize}

\usepackage{fancyhdr}

\usepackage{lscape}
\usepackage{indentfirst}
\usepackage{latexsym}
\usepackage{multirow}
\usepackage{tabls}
\usepackage{wrapfig}
\usepackage{slashbox}
\usepackage{longtable}

\bibpunct{(}{)}{;}{a}{}{,}

\shorttitle{Paper 4} 
\shortauthors{Popescu et al.}

\begin{document}

\pagestyle{fancy}

\fancyhead{}
\fancyhf{}


\fancyhead[LE,RO]{\small{\bfseries\thepage}}
\fancyhead[CE]{{\small {\bf Bogdan Popescu}, {\bf M.M. Hanson} and {\bf Bruce G. Elmegreen} (2012)}}
\fancyhead[CO]{{\small Age and Mass for 920 LMC Clusters Derived from 100 Million Monte Carlo Simulations}}

\renewcommand{\thefootnote}{\fnsymbol{footnote}}

\title{
\vspace {0.5cm} 
\LARGE{Age and Mass for 920 LMC Clusters Derived from 100 Million Monte Carlo Simulations}}
\normalsize


\author{{\bf Bogdan Popescu}\altaffilmark{1}\footnote{E-mail: popescb@mail.uc.edu}, {\bf M.M. Hanson}\altaffilmark{1}\footnote{E-mail: margaret.hanson@uc.edu} and {\bf Bruce G. Elmegreen}\altaffilmark{2}\footnote{E-mail: bge@us.ibm.com}}
\affil{$^{1}$Department of Physics, University of Cincinnati, PO Box 210011, Cincinnati, OH 45221-0011}
\affil{$^{2}$IBM Research Division, T. J. Watson Research Center, 1101 Kitchawan Road, Yorktown Heights, NY 10598}

\begin{abstract}

We present new age and mass estimates for 920 stellar clusters in the Large Magellanic Cloud (LMC) based on previously published broad-band photometry and the stellar cluster analysis package, MASSCLEAN{\fontfamily{ptm}\selectfont \textit{age}}.  Expressed in the generic fitting formula, $d^{2}N/dM dt \propto M^{\alpha}t^{\beta}$, the distribution of observed clusters is described by $\alpha = -1.5$ to $-1.6$ and $\beta = -2.1$ to $-2.2$.  For 288 of these clusters, ages have recently been determined based on stellar photometric color-magnitude diagrams, allowing us to gauge the confidence of our ages.  The results look very promising, opening up the possibility that this sample of 920 clusters, with reliable and consistent age, mass and photometric measures, might be used to constrain important characteristics about the stellar cluster population in the LMC.  We also investigate a traditional age determination method that uses a $\chi^2$ minimization routine to fit observed cluster colors to standard {\it infinite mass limit} simple stellar population models.  This reveals serious defects in the derived cluster age distribution using this method. The traditional $\chi^{2}$ minimization method, due to the variation of $U,B,V,R$ colors, will always produce an overdensity of younger and older clusters, with an underdensity of clusters in the $log(age/yr)=[7.0,7.5]$ range. Finally, we present a unique simulation aimed at illustrating and constraining the fading limit in observed cluster distributions that includes the complex effects of stochastic variations in the observed properties of stellar clusters.





\end{abstract}

\keywords{galaxies: clusters: general --- methods: analytical --- open clusters and associations: general}

\section{Introduction}

The Large Magellanic Cloud (LMC) provides a clear view of an ample number of stellar clusters with a broad age and mass range. Moreover, its well-constrained, proximal distance makes it an ideal galaxy to base observational investigations to constrain fundamental properties of stellar clusters.  The properties of interest include the mass function of stellar clusters (CMF), the rate at which star clusters form and disrupt, and the possible dependence of these rates on cluster mass and environment.  To derive these rates, one needs to know age and mass for a very large sample of stellar clusters and fully recognize and correct for incompleteness and selection bias leading to the observed sample.   Even with a very large sample, it is challenging to constrain these sought after properties.  The CMF, formation and disruption rate, along with a host of other difficult to constrain characteristics (mass-loss in stars, dynamical relaxation, tidal forces, e.g., \citeauthor*{lamers2010} \citeyear{lamers2010}) will manifest in the observed cluster sample and are not mutually independent.  Rather significant degeneracies must be addressed to independently derive the fundamental properties describing the birth, life and death of stellar clusters.
 
Because the LMC is so tantalizingly close, most of its clusters are at least partially resolved with modern telescopes. This allows for methods that use the location of individual stars populating a color-magnitude diagram (CMD) to derive the cluster age.  However, the number of clusters for which consistent, modern CMD measured ages exist has until recently remained small.  These studies did not sufficiently populate the mass-age stellar cluster distribution to constrain the fundamental cluster properties mentioned.  To extract fundamental properties such as the cluster formation rate (CFR) and the formation and disruption time scales, astronomers have derived age and mass for many tens to even hundreds of stellar clusters using integrated observations.  Historically, astronomers have relied on either integrated broad-band photometry or integrated spectroscopy, to achieve a large enough mass-age distribution to constrain the broad properties of stellar clusters in galaxies (e.g. \citeauthor*{hunter2003} \citeyear{hunter2003}; \citeauthor*{bik2003} \citeyear{bik2003}; \citeauthor*{deGrijs2003} \citeyear{deGrijs2003}; \citeauthor*{bast05} \citeyear{bast05}; \citeauthor*{lamers2005} \citeyear{lamers2005}; \citeauthor*{santos} \citeyear{santos}; \citeauthor*{bast12} \citeyear{bast12}).
 
\citeauthor*{massey2002} \citeyear{massey2002} presented an extensive survey of over 260,000 sources from the Large and Small Magellanic Clouds. \citeauthor*{hunter2003} \citeyear{hunter2003} used the \citeauthor*{massey2002} \citeyear{massey2002} photometry for a study of  939 stellar clusters. In this work we analyze 920 of these clusters that have $M_{V}$ and all three colors: $(U-B)_{0}$, $(B-V)_{0}$, and $(V-R)_{0}$.
 Now, \citeauthor*{glatt} \citeyear{glatt} have completed an extensive study, utilizing photometric data on the Magellanic Clouds combined from three sources (\citeauthor*{bica2008} \citeyear{bica2008}; \citeauthor*{z2002} \citeyear{z2002}, \citeyear{z2004}) to construct individual stellar CMDs.  \citeauthor*{glatt} \citeyear{glatt} derive ages based on a consistent set of modern stellar evolutionary models (\citeauthor*{padova2008} \citeyear{padova2008}; \citeauthor*{padova2010} \citeyear{padova2010}) for 1,193 LMC stellar clusters.  
In this work we analyze 288 clusters from \citeauthor*{glatt} \citeyear{glatt} that are included in the \citeauthor*{hunter2003} \citeyear{hunter2003} catalog.
 While the number of clusters with ages is certainly impressive, the mass of clusters is not determined and the target selection in the \citeauthor*{glatt} \citeyear{glatt} sample does not allow one to directly constrain the CMF or cluster formation or disruption rates.  However, the \citeauthor*{glatt} \citeyear{glatt} sample can be used to check on the age determination methods employed from large-scale integrated studies. 
 Here cluster samples can be better selected statistically to constrain properties of the clusters in general, but age and mass are far less certain (\citeauthor*{asa'd} \citeyear{asa'd}).  It is the precision and accuracy of current age determination methods from integrated photometry tested against the \citeauthor*{glatt} \citeyear{glatt} sample, that motivates this study.  Our ultimate goal for a future study is to use our own age and mass determinations to constrain the CMF as well as the cluster formation and disruption rate in the LMC.  Here, we strive only to reach the first step: to produce the most accurate age and mass measures for the largest sample of LMC clusters presently possible.
 
We begin this paper by drawing on the $UBVR$ photometric survey of the LMC presented by \citeauthor*{massey2002} \citeyear{massey2002} and first used to analyze LMC stellar clusters by \citeauthor*{hunter2003} \citeyear{hunter2003}.  In Section 2, ages and masses for 920 LMC clusters are derived using our own stellar cluster simulation software, MASSCLEAN, and its cluster age determination subroutine, MASSCLEAN{\fontfamily{ptm}\selectfont \textit{age}} applied to the Massey photometry.   We compare the MASSCLEAN{\fontfamily{ptm}\selectfont \textit{age}} results with traditional photometric age determination methods and measure the confidence in our new age determination method based on a comparison to the CMD ages given by \citeauthor*{glatt} \citeyear{glatt}, in Section 3. In Section 4, we briefly visit the complex issue of fading limits within stellar cluster surveys, illustrated with MASSCLEAN simulations.  This is critical to our future study to derive the cluster mass function and the destruction time scales for the LMC star clusters. Concluding remarks are given in Section 5.

\section{MASSCLEAN{\fontfamily{ptm}\selectfont \textit{age}} and {\it 100 Million} Monte Carlo Simulations}

\renewcommand{\thefootnote}{\arabic{footnote}}
\setcounter{footnote}{0}

Using the MASSCLEAN\footnote{\url{http://www.physics.uc.edu/\textasciitilde popescu/massclean/}\\ {\bf MASS}ive {\bf CL}uster {\bf E}volution and {\bf AN}alysis package is publicly available under GNU General Public License (\copyright 2007-2012 Bogdan Popescu and Margaret Hanson).} package (\citeauthor*{paper1} \citeyear{paper1}) we have built a database of integrated colors and magnitudes of stellar clusters, MASSCLEAN{\fontfamily{ptm}\selectfont \textit{colors}} (\citeauthor*{paper2} \citeyear{paper2}, \citeyear{paper3}). The traditional codes modeling Simple Stellar Populations (SSP) can only provide the integrated colors for a stellar system assuming a fully sampled stellar mass function.  A fully sampled stellar mass function will only occur when the stellar system is of very high mass, $M > 10^6 M_{\odot}$, what we will term, {\it the infinite mass limit} (e.g. \citeauthor*{lancon2000} \citeyear{lancon2000}, \citeyear{lancon2002}; \citeauthor*{lancon2010} \citeyear{lancon2010}; \citeauthor*{paper2} \citeyear{paper2}, \citeyear{paper3}; \citeauthor*{fouesneau2} \citeyear{fouesneau2}; \citeauthor*{esteban} \citeyear{esteban}).  This mass is easily achieved in single galaxies, but such a mass is rarely achieved with stellar clusters, particularly in normal galaxies.  

The MASSCLEAN models allow for a more realistic representation, recognizing the finite mass of typical stellar clusters.  MASSCLEAN allows for the stochastic fluctuations that will be observed in the {\it stellar} mass function of typical stellar clusters, and determines the expected integrated colors as a function of stellar cluster mass (\citeauthor*{paper1} \citeyear{paper1}, \citeyear{paper2}, \citeyear{paper3}). When models assuming the {\it infinite mass limit} are applied to typical stellar clusters, the mass distribution is described by fractional stars at the high end of the IMF. Since the number of stars in a real cluster is an integer, the presence or absence of massive stars in the distribution will generate fluctuations in integrated colors, both in the blue and red sides, away from the expected model colors.

In \citeauthor*{paper2} \citeyear{paper2} we showed what has long been recognized: the dispersion of stellar cluster integrated colors and magnitudes increases as the cluster mass decreases.  However, we further showed in \citeauthor*{paper3} \citeyear{paper3} the sometimes extreme, non-Guassian distribution of integrated colors and magnitudes predicted for stellar clusters, particularly for clusters with mass $< 10^4 M_{\Sun}$.   This, we will show, has severe ramifications for traditional methods used for age determination for stellar clusters with broad-band, integrated photometry. 

In \citeauthor*{paper3} \citeyear{paper3} we demonstrated the necessity of solving simultaneously for mass and age in order to reduce degeneracies in cluster characteristics derived via integrated colors. We also presented the newest addition to the MASSCLEAN package, MASSCLEAN{\fontfamily{ptm}\selectfont \textit{age}}, which uses the MASSCLEAN{\fontfamily{ptm}\selectfont \textit{colors}} database to simultaneously determine the age and mass of stellar clusters from integrated photometry. The ages determined by our program were in good agreement with the spectroscopic ages for 7 LMC clusters from \citeauthor*{santos} \citeyear{santos} (only 7 clusters from the \citeauthor*{santos} \citeyear{santos} study are found in the \citeauthor*{hunter2003} \citeyear{hunter2003} catalog).  

We tested our derived ages still further by selecting 30 clusters from the \citeauthor*{hunter2003} \citeyear{hunter2003} catalog which covered a wide range of $(U-B)_{0}$ and $(B-V)_{0}$ colors, but also had colors close to the predicted colors from traditional SSP models computed in the {\it infinite mass limit}.  In this case, the age determination based on classical fits to traditional SSP models is expected to be relatively accurate (e.g.  \citeauthor*{lancon2000} \citeyear{lancon2000}; \citeauthor*{lancon2010} \citeyear{lancon2010}; \citeauthor*{esteban} \citeyear{esteban}; \citeauthor*{paper1} \citeyear{paper1}, \citeyear{paper2}). Our MASSCLEAN{\fontfamily{ptm}\selectfont \textit{age}} results based on $U,B,V$ integrated photometry were also in good agreement with the ages from \citeauthor*{hunter2003} \citeyear{hunter2003} for this sample of 30 clusters. 

However, now we wish to investigate our MASSCLEAN{\fontfamily{ptm}\selectfont \textit{age}} predictions on considerably more LMC clusters.  In particular, we wish to investigate the reliability of MASSCLEAN{\fontfamily{ptm}\selectfont \textit{age}} predictions for clusters that are {\sl not so well behaved}.  These are clusters that lie far from the predicted colors from traditional models assuming the {\it infinite mass limit}.  We will examine the reliability of using MASSCLEAN{\fontfamily{ptm}\selectfont \textit{age}} in this difficult-to-age, color domain for stellar clusters.

\subsection{Masses and ages of 920 LMC Clusters using MASSCLEAN{\fontfamily{ptm}\selectfont \textit{age}}}

The ages derived for this study use the newest version of the MASSCLEAN{\fontfamily{ptm}\selectfont \textit{colors}} database, now based on over $100$ million Monte Carlo stellar cluster simulations and using Padova isochrones (\citeauthor*{padova2008} \citeyear{padova2008}, Girardi et al. 2010) with $Z=.008$ metalicity. The simulations were done using a Kroupa IMF (\citeauthor*{Kroupa2002} \citeyear{Kroupa2002}) with $0.1$ $M_{\Sun}$ and $120$ $M_{\Sun}$ mass limits. The age range in the database extends from $[6.6,9.5]$ in $log(age/yr)$, and includes 95 mass intervals over the mass range $200$-$100,000$ $M_{\Sun}$.  Stellar cluster ages and masses are then computed using their observed, extinction-corrected $M_{V}$, $(U-B)_{0}$, $(B-V)_{0}$, $(V-R)_{0}$ integrated photometry from the \citeauthor*{hunter2003} \citeyear{hunter2003} catalog. The addition of the $(V-R)_{0}$ color, over what was used to derive cluster ages in \citeauthor*{paper3} \citeyear{paper3}, further helps to break degeneracies found in age and mass (see Fig.\ 10 in \citeauthor*{paper3} \citeyear{paper3}) and to better constrain the characteristics of clusters found to lie well outside the colors predicted from {traditional models assuming the {\it infinite mass limit}. 

MASSCLEAN{\fontfamily{ptm}\selectfont \textit{age}} uses the integrated photometry for each cluster to simultaneously derive the most probable cluster age and mass in the form of an age-mass probability distribution (please see \citeauthor*{paper3} \citeyear{paper3} for a full description of the routine).  In this newest study, we aspired to further increase the accuracy of our results by performing multiple runs of MASSCLEAN{\fontfamily{ptm}\selectfont \textit{age}} for each cluster, using the integrated photometry perturbed over the range of the photometric errors from \citeauthor*{hunter2003} \citeyear{hunter2003}.  A single MASSCLEAN{\fontfamily{ptm}\selectfont \textit{age}} run takes about 30 minutes on a standard quad-core desktop computer.  Multiple runs with perturbed integrated magnitudes and colors for the 920 clusters required about 8 months on the same computer. 

The age and mass results for 920 LMC clusters derived in this way using MASSCLEAN{\fontfamily{ptm}\selectfont \textit{age}} are presented in Figures \ref{fig:paper4-01} -- \ref{fig:paper4-02new}.  The entire dataset from Figures \ref{fig:paper4-01} -- \ref{fig:paper4-02new} is also presented in Tables \ref{table1} and \ref{table2}. 
A subset of 288 clusters which have CMD ages from Glatt et al. 2010 is presented in Table \ref{table1}. The CMD age is presented in Column 7, and $E(B-V)$ in Column 8. The remaining 632 clusters are presented in Table \ref{table2}.
 The data from the \citeauthor*{hunter2003} \citeyear{hunter2003} catalog are shown in the Columns 2--5 in both tables. The MASSCLEAN{\fontfamily{ptm}\selectfont \textit{age}} results are presented in the Columns 9--10 in the Table \ref{table1}, and in the columns 7--8 in the Table \ref{table2}.

\begin{figure}[htp]
\centering
\includegraphics[angle=270,width=0.5\textwidth, bb= 100 100 520 770]{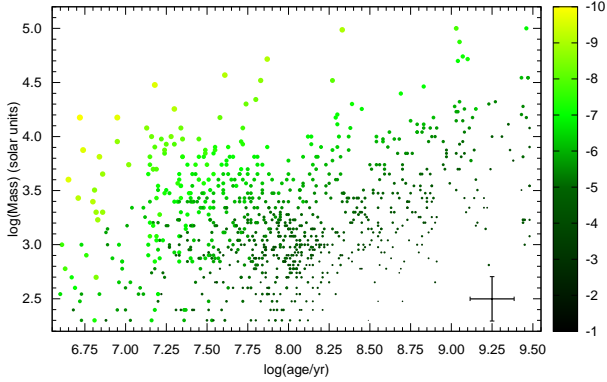}
\caption[]{\footnotesize The age and mass for 920 LMC clusters, displayed on the $log(M/M_{\Sun})$ vs. $log(age/yr)$ plot.  The dots are color-coded based on the absolute magnitude in $V$ Band, $M_{V}$. The size of the dots are further scaled to correspond with cluster $M_{V}$.  Each point has its own unique error (listed in Tables \ref{table1} and \ref{table2}), but the mean error is given in the lower right corner. \normalsize}\label{fig:paper4-01}
\end{figure}


Figure \ref{fig:paper4-01} displays the MASSCLEAN values for cluster mass and age as $log(M/M_{\Sun})$ vs. $log(age/yr)$.   The mean error in mass and age is shown in the figure, though each cluster has its own uniquely derived age and mass error, as given in Tables \ref{table1} and \ref{table2} (columns 9 \& 10, and 7 \& 8, respectively).  Figure \ref{fig:paper4-01} is color-coded to show the clusters integrated $M_{V}$ magnitude. The size of the dots is further scaled with $M_{V}$, more luminous clusters being represented by larger dots.  What is immediately clear from this figure is that the LMC cluster sample is dominated by low-mass clusters, with masses typically well below $10^{4} M_{\Sun}$. This result agrees with the recent mass estimation for the cluster population of the LMC (e.g. \citeauthor*{pessev2008} \citeyear{pessev2008}; \citeauthor*{chandar2010a} \citeyear{chandar2010a}; \citeauthor*{chandar2010b} \citeyear{chandar2010b}; \citeauthor*{larsen2010} \citeyear{larsen2010}).  One also immediately sees the decreased luminosity with age for clusters of identical mass.

In Figure \ref{fig:paper4-10new} is shown the entire $UBV$ data set of 920 clusters, taken from \citeauthor*{massey2002} \citeyearpar{massey2002}.  \citeauthor*{glatt} \citeyear{glatt} extinction values have been applied when available, otherwise, a mean extinction correction of $E(B-V) = 0.13$, as used by \citeauthor*{hunter2003} \citeyear{hunter2003}, was applied.  The clusters are seen to lie over a large range of colors, though mostly scattered about the traditional {\it infinite mass limit} prediction (the continuous line). In the left panel,  blue, negative colors belong to young clusters, smoothly transition to older, red clusters, in the lower right portion of the diagram.  In the right panel of Figure \ref{fig:paper4-10new}, the traditional {\it infinite mass limit} line is colored to show the age.  For the most part, the ages assigned by MASSCLEAN{\fontfamily{ptm}\selectfont \textit{age}} follow along the colors predicted from the traditional SSP models computed in the {\it infinite mass limit}.  However, there are a number of clusters who's color (and thus age) is not consistent with its neighbors or the {\it infinite mass limit} line.  Their mass, illustrated in the right panel by point size, is typically very low.

\begin{figure*}[htp]
\centering
\includegraphics[angle=0,width=0.49\textwidth, bb= 65 120 557 670]{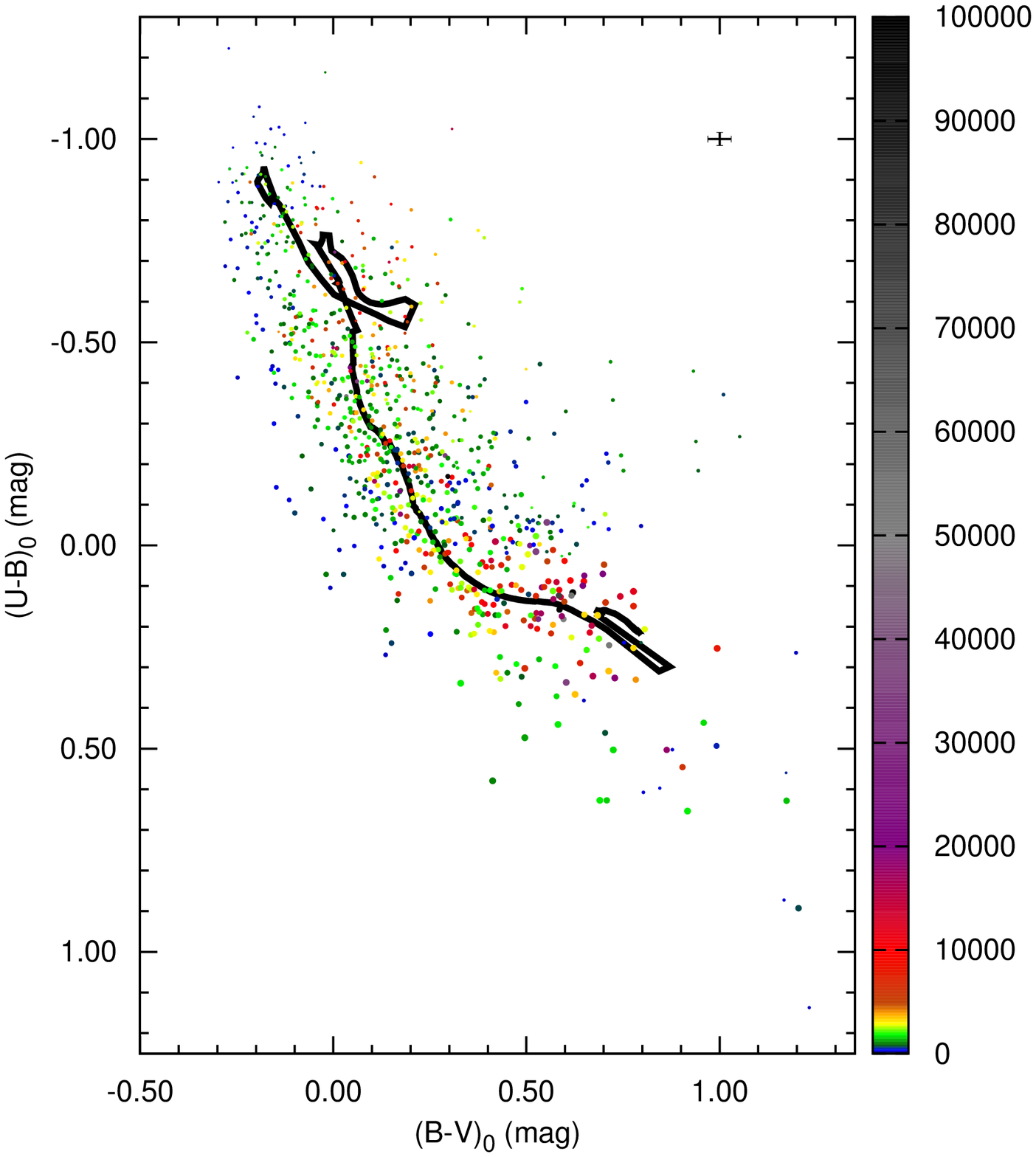}
\includegraphics[angle=0,width=0.49\textwidth, bb=  65 120 557 670]{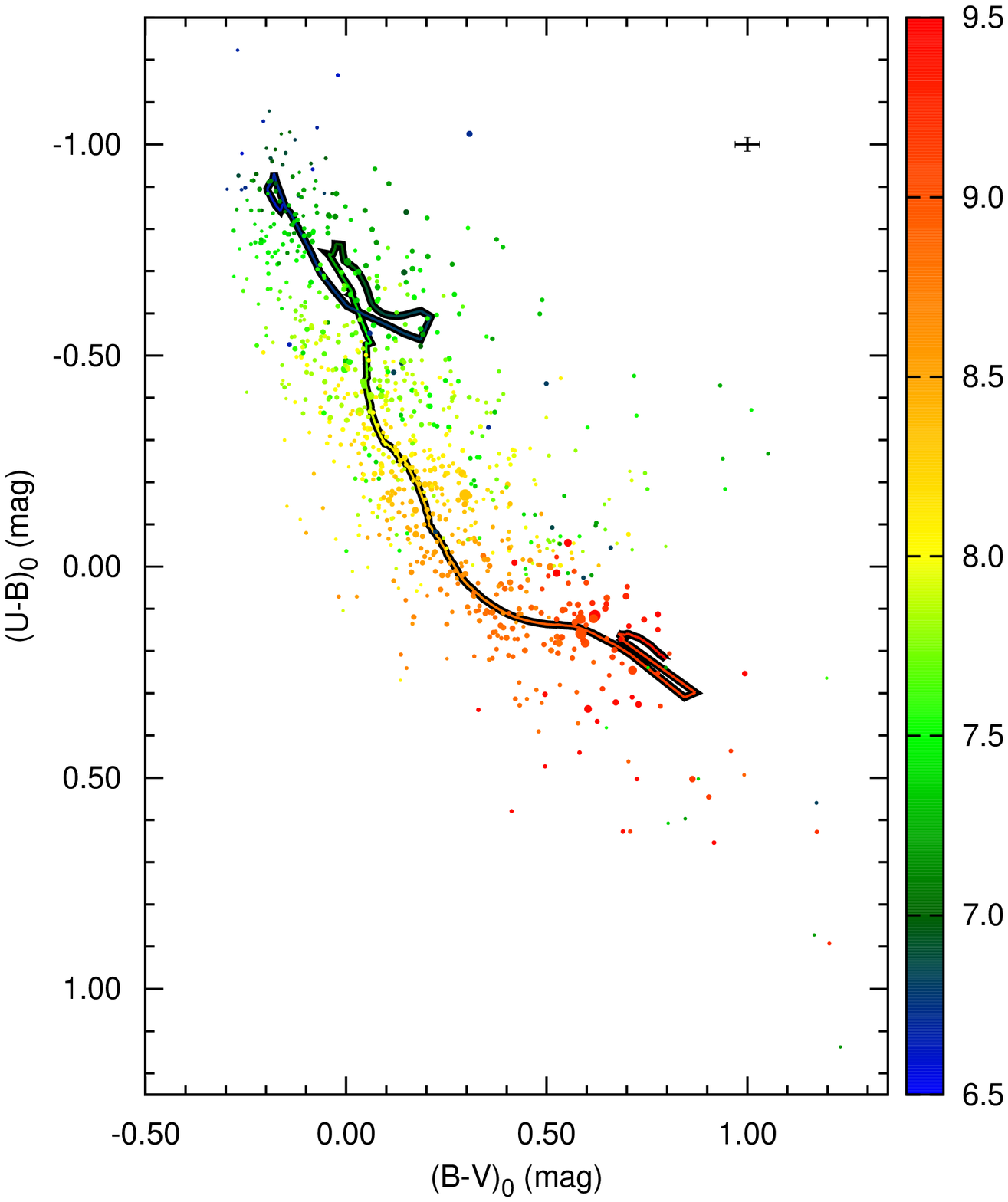}

\vspace{-0.4cm}
\includegraphics[angle=270,width=6.4cm, bb= 285 144 330 680]{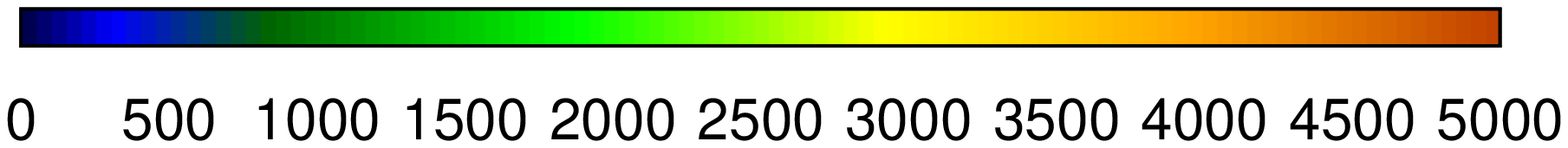}
\hspace{8.6 cm}

\caption{\footnotesize {\it Left:} LMC clusters on $(U-B)_{0}$ vs. $(B-V)_{0}$ color-color diagram. The color is scaled with the mass, and the size of the dots is scaled with the age. The traditional SSP model, assuming an infinite-mass system is represented in the black line for eye-guiding. The color scale presented bellow the plot magnifies the $0$$-$$5,000$ $M_{\Sun}$ range.  {\it Right:} LMC clusters on $(U-B)_{0}$ vs. $(B-V)_{0}$ color-color diagram. The color is scaled with the age, and the size of the dots is scaled with the mass. The SSP predictions, assuming the limit of an infinite-mass system is also colored to display the age. The integrated colors are from \citeauthor*{hunter2003} \citeyear{hunter2003} catalog and have been corrected for extinction.  Masses and ages given are computed using MASSCLEAN{\fontfamily{ptm}\selectfont \textit{age}}. Each point has its own unique error (listed in Tables \ref{table1} and \ref{table2}), but the mean error is given in the upper right corner. \normalsize}\label{fig:paper4-10new}
\end{figure*}

The dispersion of observed cluster colors, away from the traditional SSP {\it infinite mass limit} and as a function of mass, is illustrated in Figure \ref{fig:paper4-05old}. Here we have plotted the difference,$\Delta$(U-B)$_0$, between the traditional, mass-insensitive, {\it infinite mass limit} color predicted for that age and the clusters observed color versus the MASSCLEAN{\fontfamily{ptm}\selectfont \textit{age}} derived cluster mass.  Once the stellar cluster mass exceeds 10,000 $M_{\Sun}$, the dispersion is small enough to be nearly dominated by photometric error in most cases.  But below this mass, the dispersion is very real, and is due to the very poor sampling of the stellar mass function.  The poorly sampled stellar mass function leads to wildly varying observed colors as compared to those predicted by traditional models computed in the {\it infinite mass limit}. 

\begin{figure}[htp]
\centering
\includegraphics[angle=270,width=8.5cm, bb= 115 90 515 735]{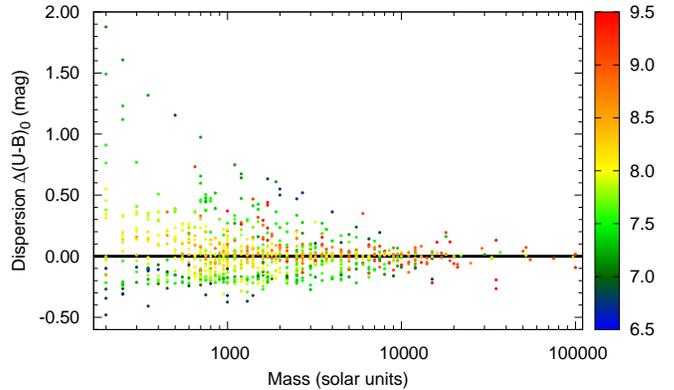}
\caption[]{\footnotesize The dispersion, $(U-B)_{0}$, shows the difference between the traditional SSP predicted color and the observed cluster's color, plotted versus the cluster mass. The dots are also color-coded to display the age. Low mass clusters show the largest dispersion, though older clusters (seen as yellow and red), even with low mass, do fall reasonably close to the predicted SSP colors.  \normalsize}\label{fig:paper4-05old}
\end{figure}

\subsection{Stellar Clusters Fade in Number and Luminosity with Time}

Yet another representation of the LMC cluster sample is given in Figure \ref{fig:paper4-02new}.   Here, we show the integrated magnitudes $M_{V}$ vs. $log(age/yr)$.   A few things are readily apparent.  First, the number of clusters in the sample as a function of time decreases.  Ignoring the colors for now, the density of dots looks relatively constant, but with logarithmic age bins this suggests a factor of ten drop in the number of clusters for each factor of ten increase in age.  This has been noted already in studies of clusters of low mass, 100 - 1000 M$_{\Sun}$ (\citeauthor*{lada} \citeyear{lada}) all the way up to clusters of very high mass, $10^4$ - $10^6$ M$_{\Sun}$ (\citeauthor*{fall2009} \citeyear{fall2009}). 

\begin{figure}[htp]
\centering
\includegraphics[angle=270,width=0.5\textwidth, bb= 100 100 520 770]{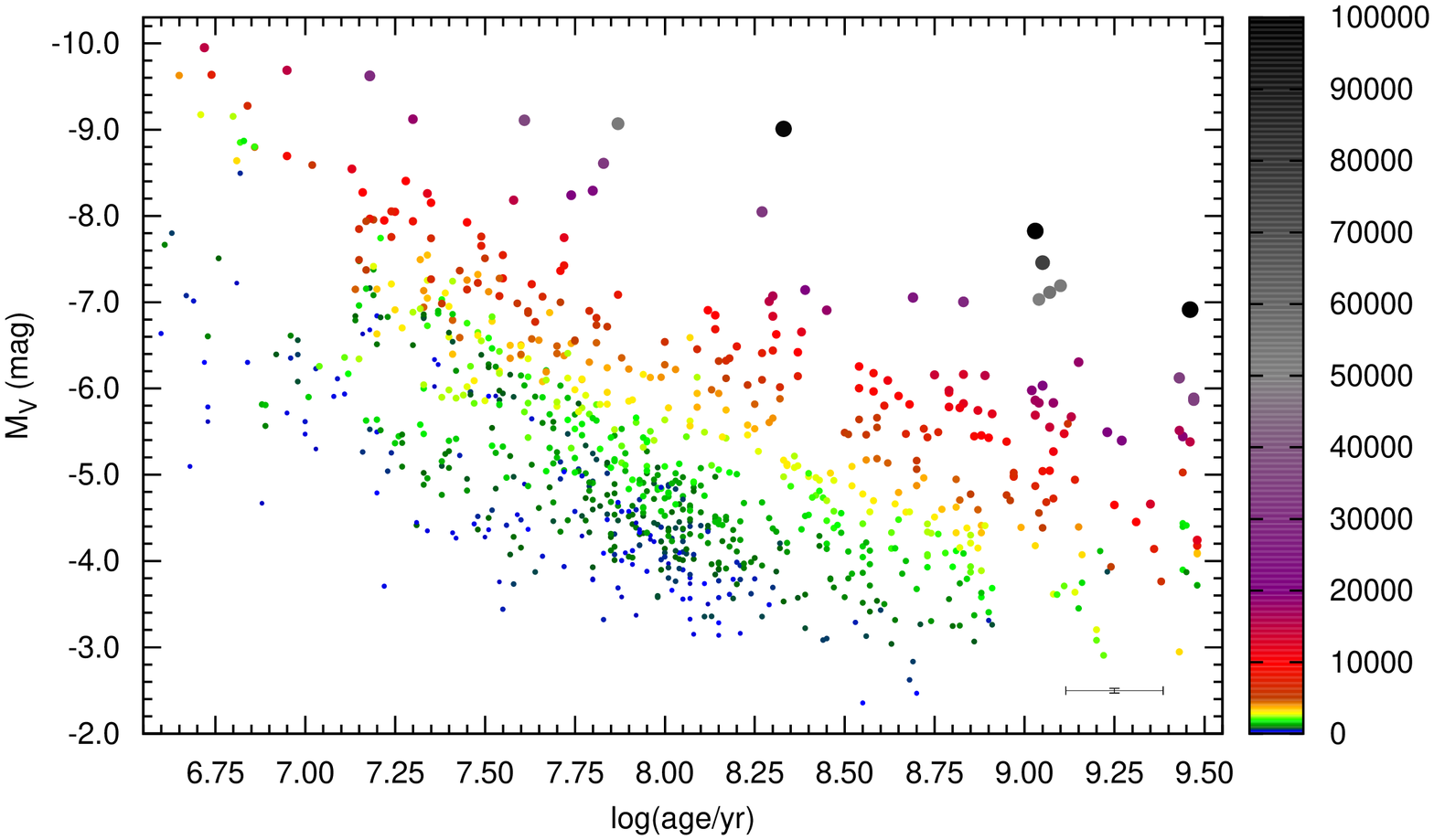}

\includegraphics[angle=270,width=0.395\textwidth, bb= 285 144 330 680]{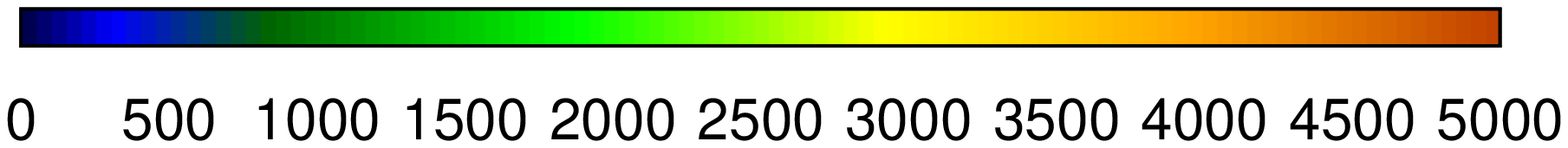}
\caption[]{\footnotesize Integrated absolute magnitude in $V$ Band, $M_{V}$, vs. MASSCLEAN-derived $log(age/yr)$ for 920 LMC clusters. The integrated magnitudes are from \citeauthor*{hunter2003} \citeyear{hunter2003} catalog.  Dot color and size is further scaled with MASSCLEAN-derived mass. The color scale presented bellow the plot magnifies the $0$$-$$5,000$ $M_{\Sun}$ range. The mean error is given in the lower right corner. \normalsize}\label{fig:paper4-02new}
\end{figure}

Secondly, now following the color-coding of mass with age, one sees an obvious fading of cluster luminosity with time.  And a related point, the number of high luminosity clusters is diminishing rather quickly with time.  For the LMC sample, the colors (representing clusters of similar mass) seem to stay fairly segregated, following a reasonably smooth function of decreased luminosity with time among the highest mass clusters.  However, this segregation begins to break down below masses of a few thousand $M_{\Sun}$, where colors (representing masses) begin to overlap with each other.  This is due to the increasing effect of stochastic fluctuations in the IMF for the low-mass clusters.  In such clusters, an individual star at the top of the current mass function (or lack there of) can cause rather large deviations in the clusters observed total magnitude (e.g. \citeauthor*{lancon2010} \citeyear{lancon2010};
\citeauthor*{fouesneau2} \citeyear{fouesneau2}; \citeauthor*{lancon2011} \citeyear{lancon2011};  \citeauthor*{paper3} \citeyear{paper3}; \citeauthor*{esteban} \citeyear{esteban}; \citeauthor*{slug} \citeyear{slug}).

\subsection{Comparison with CMD Ages}

A subset of 288 clusters from the \citeauthor*{hunter2003} \citeyear{hunter2003} catalog were also studied by \citeauthor*{glatt} \citeyear{glatt} where they derived CMD ages.  The overlapping set of clusters between these two studies are listed in Table \ref{table1}. We used MASSCLEAN{\fontfamily{ptm}\selectfont \textit{age}} to determine the mass and age of these clusters using the \citeauthor*{hunter2003} \citeyear{hunter2003} integrated magnitudes and colors, corrected for the individual extinction given for each cluster by \citeauthor*{glatt} \citeyear{glatt} with the CCM (\citeauthor*{ccm} \citeyear{ccm}) extinction law. 

The MASSCLEAN{\fontfamily{ptm}\selectfont \textit{age}} results for this subset of 288 clusters are presented in Figure \ref{fig:paper4-03new} as a $log(M/M_{\Sun})$. vs. $log(age/yr)$ plot, similar to Figure \ref{fig:paper4-01}.   Again, the dots are colors coded to show $M_{V}$, and their size is scaled with $M_{V}$.  The \citeauthor*{glatt} \citeyear{glatt} study discarded clusters showing ages over 1 Gyr.  Also, because they selected clusters that had been previously identified by \citeauthor*{bica2008} \citeyear{bica2008} as clusters without any associated emission or HII regions, they do not expect clusters younger than 10 Myr.   MASSCLEAN{\fontfamily{ptm}\selectfont \textit{age}} did however derive solutions for the \citeauthor*{glatt} \citeyear{glatt} clusters that were outside their estimated age range for the sample (Figure \ref{fig:paper4-03new}), $log(age/yr)=[7.0,9.0]$.

\begin{figure}[htp]
\centering
\includegraphics[angle=270,width=0.5\textwidth, bb= 100 100 520 770]{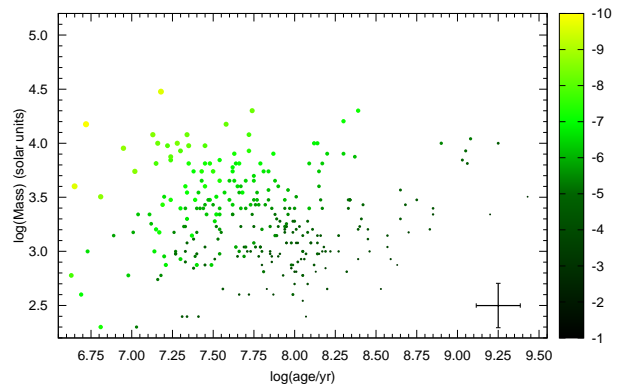}
\caption[]{\footnotesize The MASSCLEAN-derived age and mass for 288 LMC clusters with CMD age from \citeauthor*{glatt} \citeyear{glatt}, displayed on the $log(M/M_{\Sun})$ vs. $log(age/yr)$ plot.  Similar to Figure \ref{fig:paper4-01}, the dots are color-coded based on $M_{V}$ and the dot size is scaled with cluster $M_{V}$. The mean error is given in the lower right corner. \normalsize}\label{fig:paper4-03new}
\end{figure}

How do our MASSCLEAN{\fontfamily{ptm}\selectfont \textit{age}} results compare to the \citeauthor*{glatt} \citeyear{glatt} CMD ages?  This is shown in Figure \ref{fig:paper4-04new}. In the upper panel the dots are color-coded to display the integrated $M_{V}$ magnitude.  Out of a total of 288 clusters, 17 where placed by MASSCLEAN{\fontfamily{ptm}\selectfont \textit{age}} outside the \citeauthor*{glatt} \citeyear{glatt} $log(age/yr)=[7.0,9.0]$ expected age range, given by the gray box in Figure \ref{fig:paper4-04new}.  Within the box, the remaining 271 clusters show reasonably good correlation between the two methods, most are within $0.135$ $\Delta \phd log(age/yr)$ (which is the average error of MASSCLEAN{\fontfamily{ptm}\selectfont \textit{age}}), represented by the parallel lines to each side of the identity line in Figure \ref{fig:paper4-04new} (see also Table \ref{table1}).  No strong bias, offset or trend is apparent, though there is a possible over-density below the line (indicating MASSCLEAN{\fontfamily{ptm}\selectfont \textit{age}} may be slightly underestimating cluster ages as compared to \citeauthor*{glatt} \citeyear{glatt}).  Cluster mass was not calculated by \citeauthor*{glatt} \citeyear{glatt}, but this is something that comes from the MASSCLEAN{\fontfamily{ptm}\selectfont \textit{age}} solution along with age.  The masses derived in this way are indicated by color-coding in the lower plot of Figure \ref{fig:paper4-04new}.  Several fairly massive clusters (shown in red-orange) were given intermediate ages by \citeauthor*{glatt} \citeyear{glatt}, but MASSCLEAN{\fontfamily{ptm}\selectfont \textit{age}} determined them to be fairly old.  

\begin{figure}[htp]
\centering
\includegraphics[angle=270,width=0.5\textwidth, bb= 115 170 525 650]{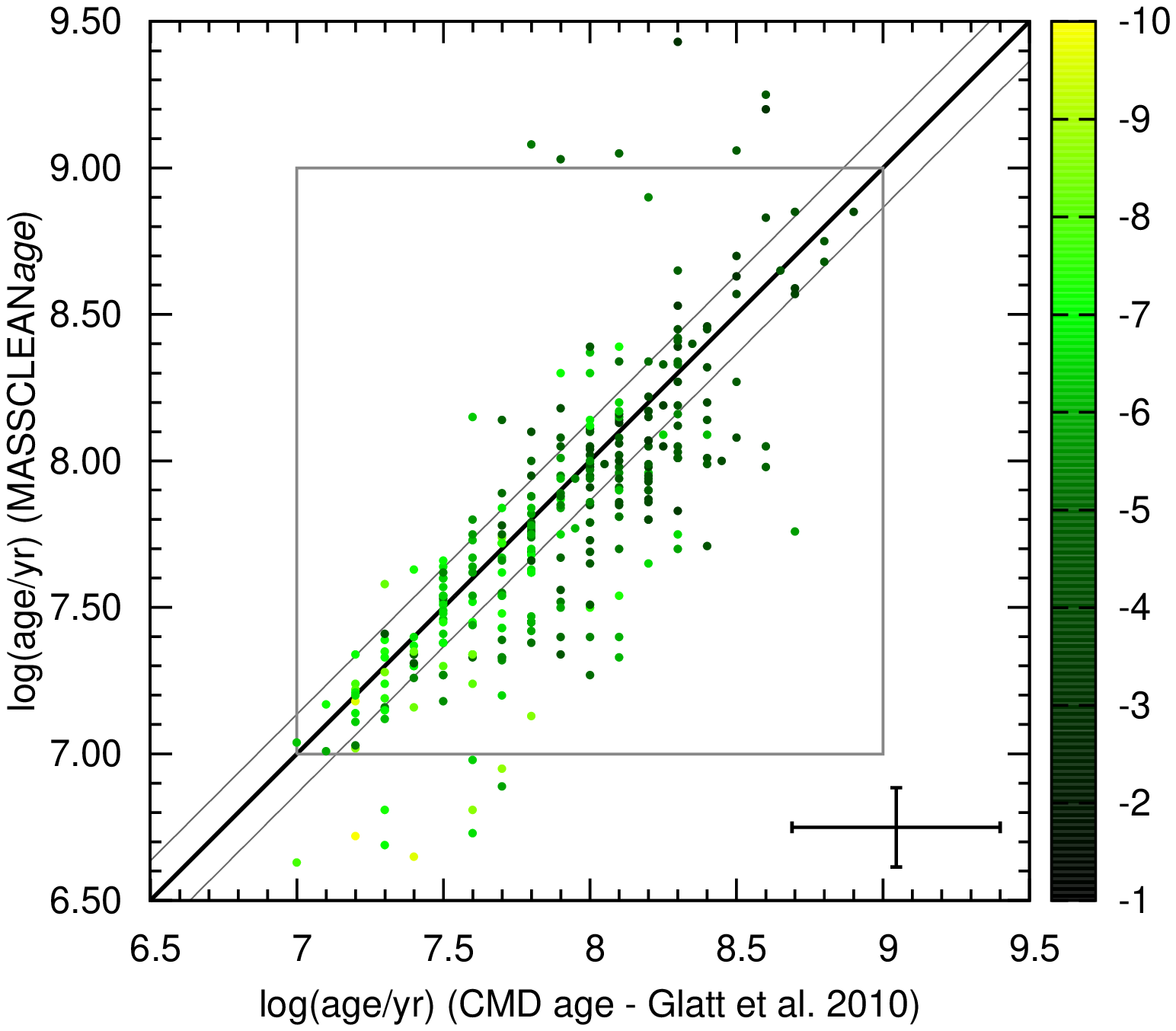}

\includegraphics[angle=270,width=0.5\textwidth, bb= 115 170 525 650]{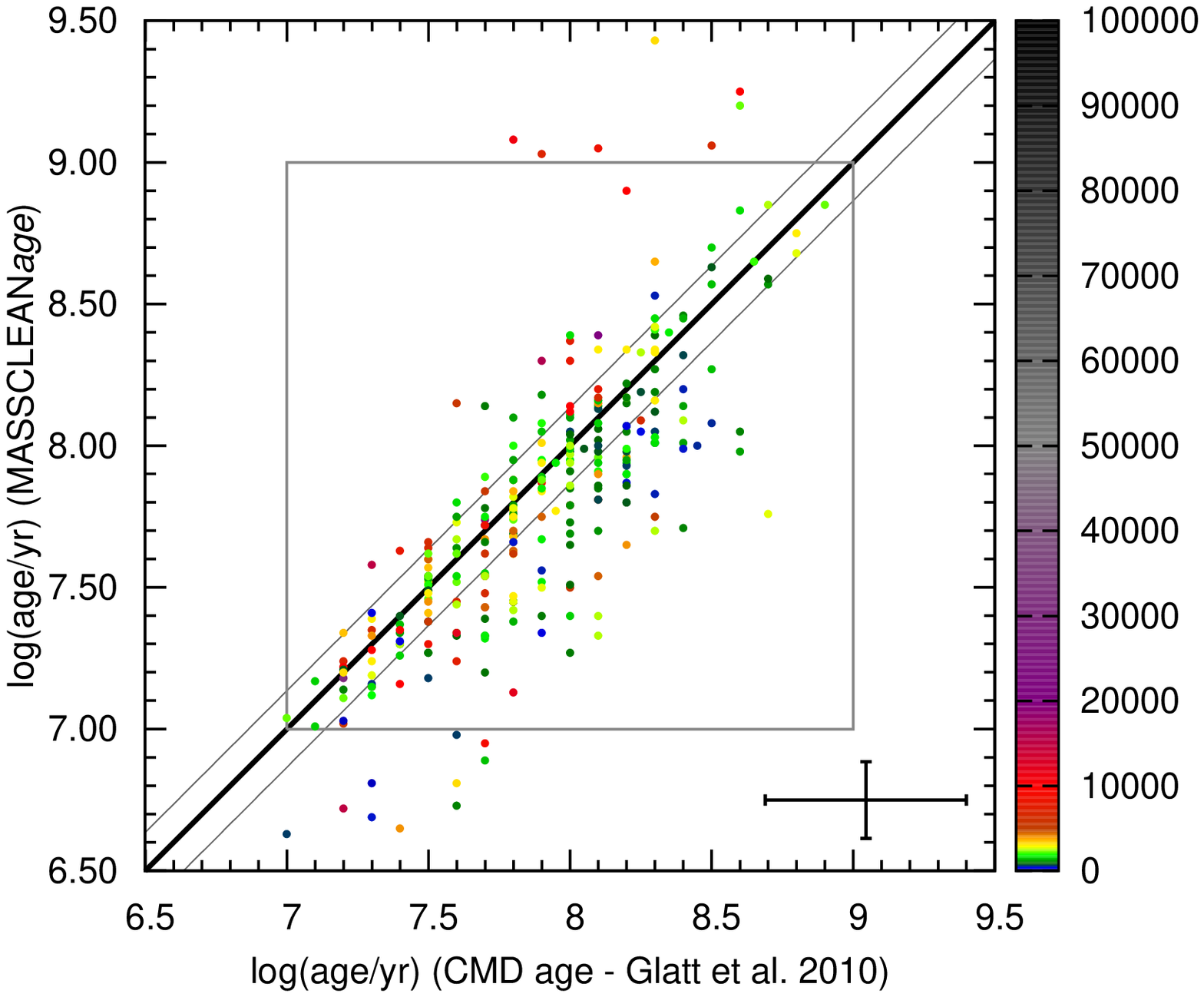}
\vspace{-0.7 cm}

\hspace{0.5 cm}
\includegraphics[angle=270,width=0.353\textwidth, bb= 280 144 330 680]{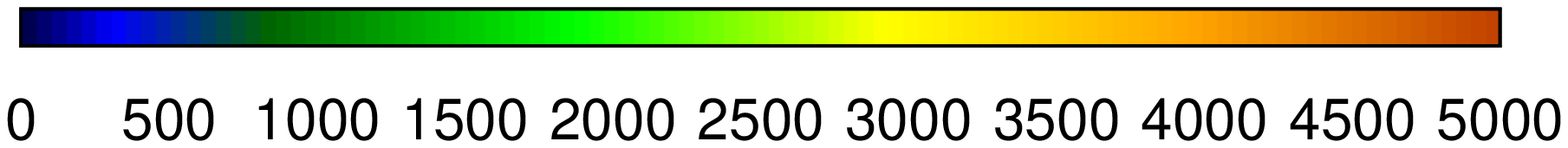}
\caption[]{\footnotesize Comparison of CMD ages from \citeauthor*{glatt} \citeyear{glatt}  with MASSCLEAN ages for 288 LMC clusters.  {\it Top:} The dots are color-coded to display the integrated $M_{V}$ magnitude. {\it Bottom:} The dots are color-coded to show the mass of the cluster derived using MASSCLEAN{\fontfamily{ptm}\selectfont \textit{age}}.  The gray box indicates the age range \citeauthor*{glatt} \citeyear{glatt} estimated for their sample. The mean error is given in the lower right corner. \normalsize}\label{fig:paper4-04new}
\end{figure}

A further comparison is given in Figure \ref{fig:paper4-05new}.   Here we repeat the plot of Figure \ref{fig:paper4-02new}, but plot only those clusters in common with the \citeauthor*{glatt} \citeyear{glatt} study.  The magnitudes all come from \citeauthor*{hunter2003} \citeyear{hunter2003}, but the ages come from MASSCLEAN{\fontfamily{ptm}\selectfont \textit{age}} on the top panel, and the CMD ages derived from \citeauthor*{glatt} \citeyear{glatt} on the bottom panel (hence the quantized nature of data). The masses are again color coded. In the bottom panel the mass and the integrated $M_{V}$ magnitude did not appear so well correlated like in the top panel. This is in part because the mass comes from MASSCLEAN{\fontfamily{ptm}\selectfont \textit{age}} and the age displayed is from \citeauthor*{glatt} \citeyear{glatt}.   The later ages will be be quite different from the MASSCLEAN{\fontfamily{ptm}\selectfont \textit{age}} used to derive the displayed mass, such as the 17 clusters located outside the $log(age/yr)=[7.0,9.0]$ age range of \citeauthor*{glatt} \citeyear{glatt}.  Also, while \citeauthor*{glatt} \citeyear{glatt} provide a large and consistent set of ages for LMC clusters, the age step used in age determination is quite large ($\Delta \phd log(age/yr)=0.05-0.10$).  The age error bars are also quit large (the mean age error is displayed in Figures \ref{fig:paper4-04new} and \ref{fig:paper4-05new}, and the values for each cluster are presented in Table \ref{table2}).

\begin{figure}[htp]
\centering
\includegraphics[angle=270,width=0.5\textwidth, bb= 100 100 520 770]{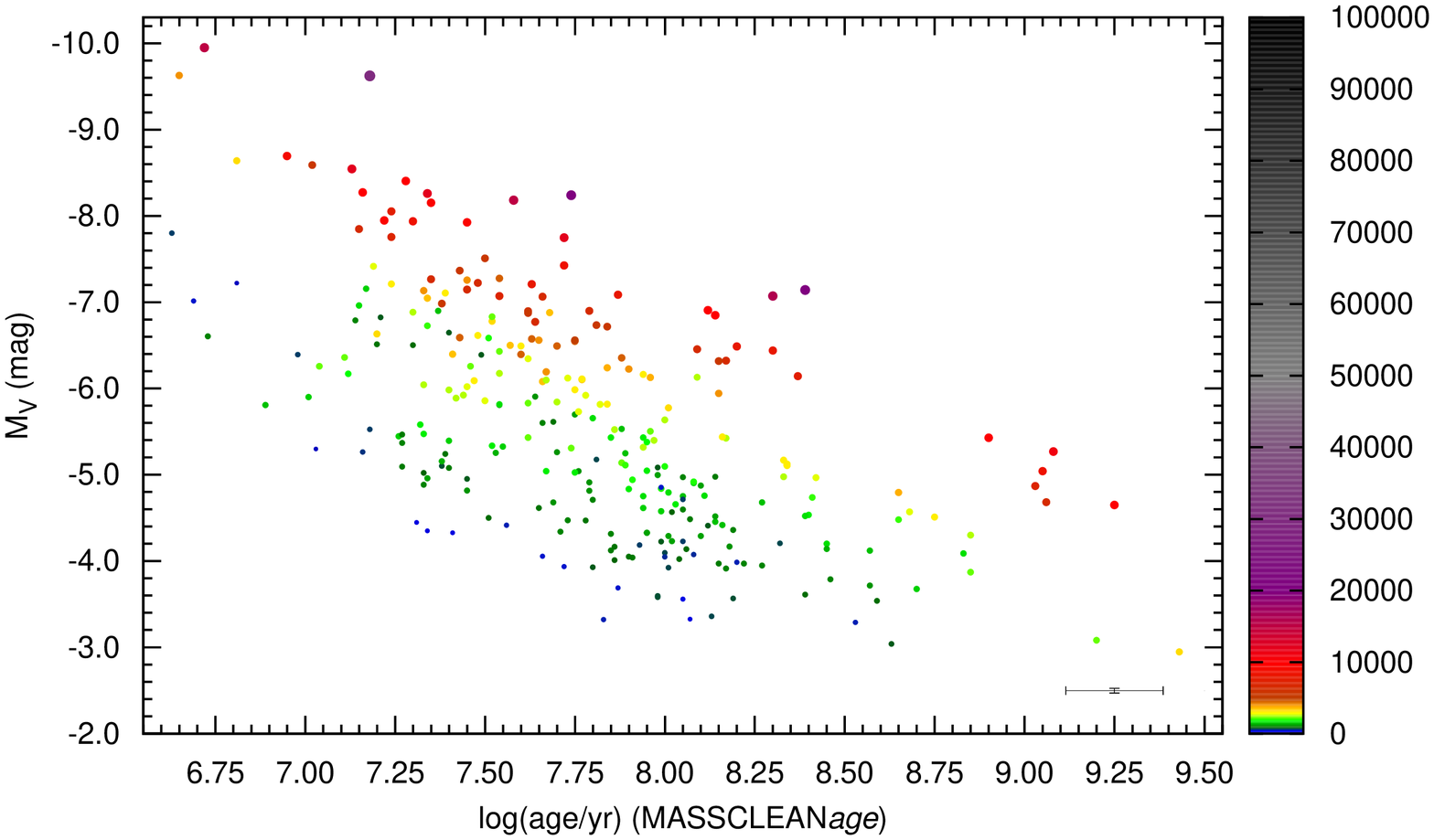}

\vspace{-0.15 cm}
\includegraphics[angle=270,width=0.395\textwidth, bb= 285 144 330 680]{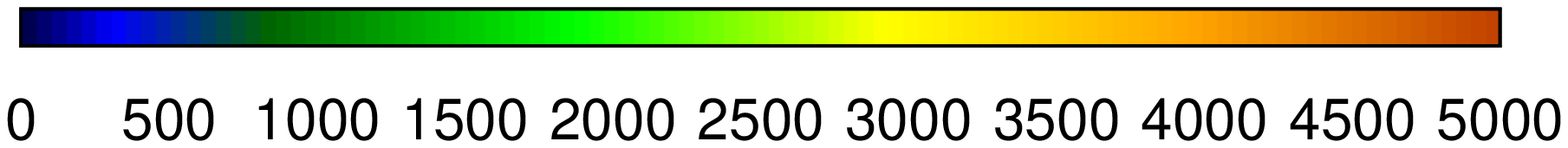}


\vspace{-0.1 cm}
\includegraphics[angle=270,width=0.5\textwidth, bb= 100 100 520 770]{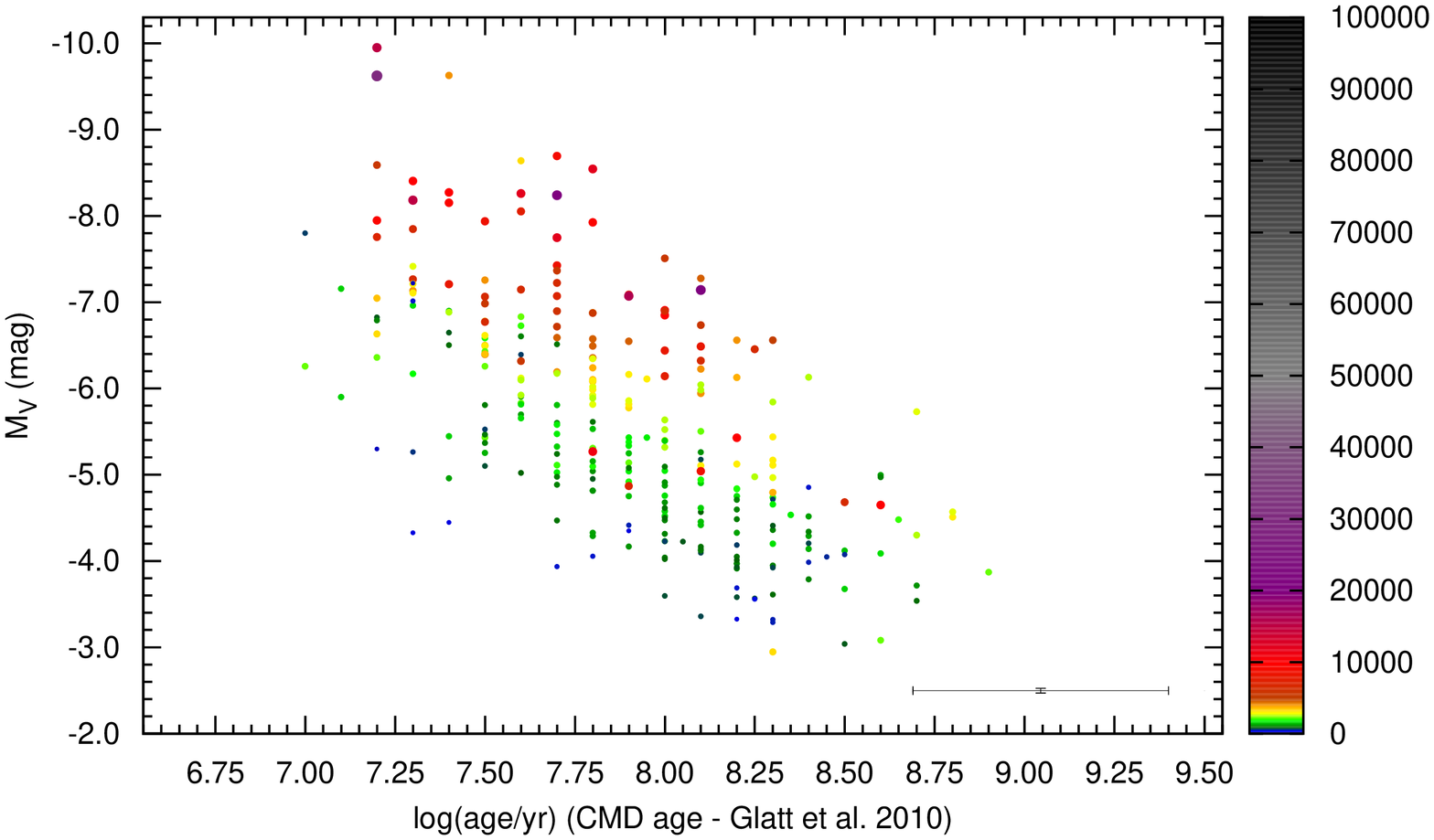}

\vspace{-0.15 cm}
\includegraphics[angle=270,width=0.395\textwidth, bb= 285 144 330 680]{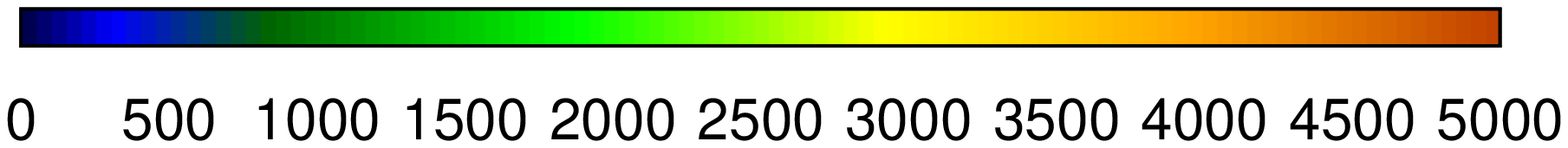}


\caption[]{\footnotesize Similar to Figure \ref{fig:paper4-02new}, integrated $M_{V}$ magnitudes vs. $log(age/yr)$ for 288 LMC clusters with CMD age. The magnitude is from \citeauthor*{hunter2003} \citeyear{hunter2003}. The dots are color-coded based on the cluster mass (derived by MASSCLEAN{\fontfamily{ptm}\selectfont \textit{age}}) and dot size is scaled to $M_{V}$.  The ages computed using MASSCLEAN{\fontfamily{ptm}\selectfont \textit{age}} are used in the {\it Top} panel, and the CMD ages from \citeauthor*{glatt} \citeyear{glatt} are used in the {\it Bottom} panel. The mean error is given in the lower right corner. \normalsize}
\label{fig:paper4-05new}
\end{figure}

\section{Comparison with Earlier Age Studies Using Integrated Magnitudes}\label{traditional}

\subsection{The \citeauthor*{hunter2003} \citeyear{hunter2003} Ages}

The first study to derive ages using the $UBVR$ colors from the sample in Tables \ref{table1} and \ref{table2} was of course the \citeauthor*{hunter2003} \citeyear{hunter2003} study.  Their ages are all fully tabulated and published, and we have listed them along side our own MASSCLEAN ages in Tables \ref{table1} and \ref{table2}.  \citeauthor*{hunter2003} \citeyear{hunter2003} used the Geneva stellar evolutionary models (\citeauthor*{geneva1} \citeyear{geneva1}) with $Z = 0.008$.  They assumed a single reddening of $E(B-V) = 0.13$ for all clusters.   For older clusters (age $>$ 1 Gyr), \citeauthor*{hunter2003} \citeyear{hunter2003} used the age-color relationship given in \citeauthor*{searle1973} \citeyear{searle1973}.   \citeauthor*{hunter2003} \citeyear{hunter2003} acknowledge the difficulty of assigning ages when clusters fail to fall in regions covered by the models.   For clusters lying significantly distant from the model predictions, they wisely chose not to assign an age.  This did represent a rather sizable fraction of the clusters in their study.  These clusters without \citeauthor*{hunter2003} \citeyear{hunter2003} ages are listed in Tables \ref{table1} and \ref{table2} as having $log(age/yr) = 10.0$.   We are grateful the \citeauthor*{hunter2003} \citeyear{hunter2003} study included the photometric properties for all clusters in their extended tables, including those without ages.  This allowed us to use MASSCLEAN{\fontfamily{ptm}\selectfont \textit{age}} to derive ages for the full sample and consider its effectiveness in deriving ages for clusters lying well outside the range of standard models (see Section 3.3). 

How do the two age determination methods compare?  We have plotted the cluster ages derived using MASSCLEAN{\fontfamily{ptm}\selectfont \textit{age}} versus those given by \citeauthor*{hunter2003} \citeyear{hunter2003} in Figure \ref{fig:paper4-06new} for the full sample of 920 clusters. In the upper panel the dots are color-coded to display the dispersion in $(U-B)_{0}$ with respect to the {\it infinite mass limit}, and in the lower panel the colors show the MASSCLEAN derived mass for clusters. The mean MASSCLEAN{\fontfamily{ptm}\selectfont \textit{age}} errors are displayed by the gray lines parallel to the identity line, similarly to Figure \ref{fig:paper4-04new}. 
A large number of clusters line up along the $log(age/yr)=10.00$ line, but these are clusters for which the authors felt they could not assign a confident age based on the models they had.  While there is in general reasonable agreement, there is a definite slope flattening seen in the figure.  MASSCLEAN{\fontfamily{ptm}\selectfont \textit{age}} has typically assigned slightly older ages for the \citeauthor*{hunter2003} \citeyear{hunter2003} clusters with ages under $log(age/yr) = 8.0$.  

There are two different effects at play here.   First, the \citeauthor*{hunter2003} \citeyear{hunter2003} study derives ages using the Geneva models (\citeauthor*{geneva1} \citeyear{geneva1}).  \citeauthor*{glatt} \citeyear{glatt}  has shown that there is a detectable offset between the Padova and Geneva isochrones, in such a way that Geneva clusters were typically several dex younger than the same cluster aged using Padova isochrones.  This occurred only for cluster with $log(age/yr)< 8.0$.  This explains some of the offset seen in Figure \ref{fig:paper4-06new} among the clusters with ages below log(age) $< 8.0$.

\begin{figure}[htp]
\centering
\includegraphics[angle=270,width=0.5\textwidth, bb= 115 170 525 650]{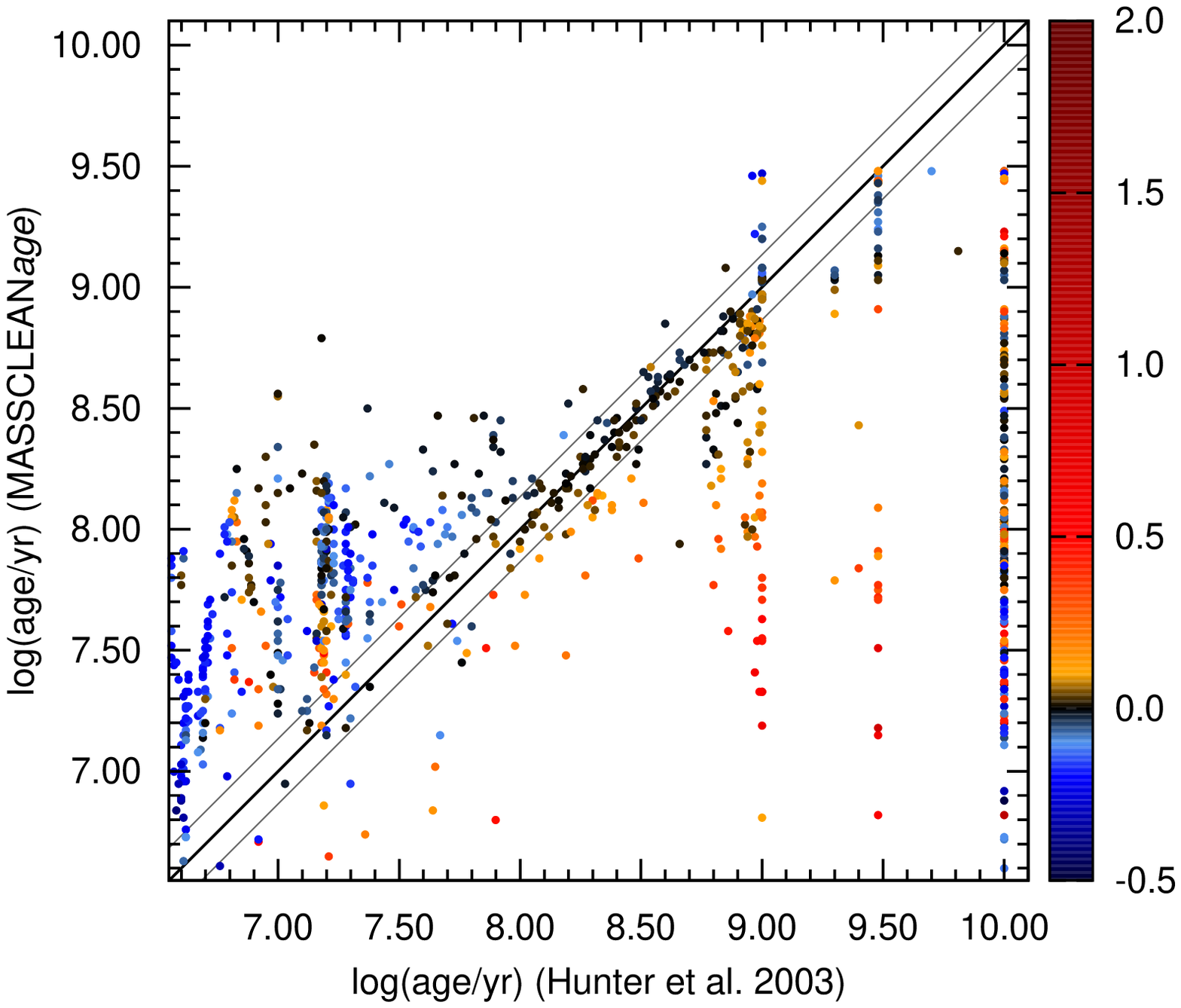}

\includegraphics[angle=270,width=0.5\textwidth, bb= 115 170 525 650]{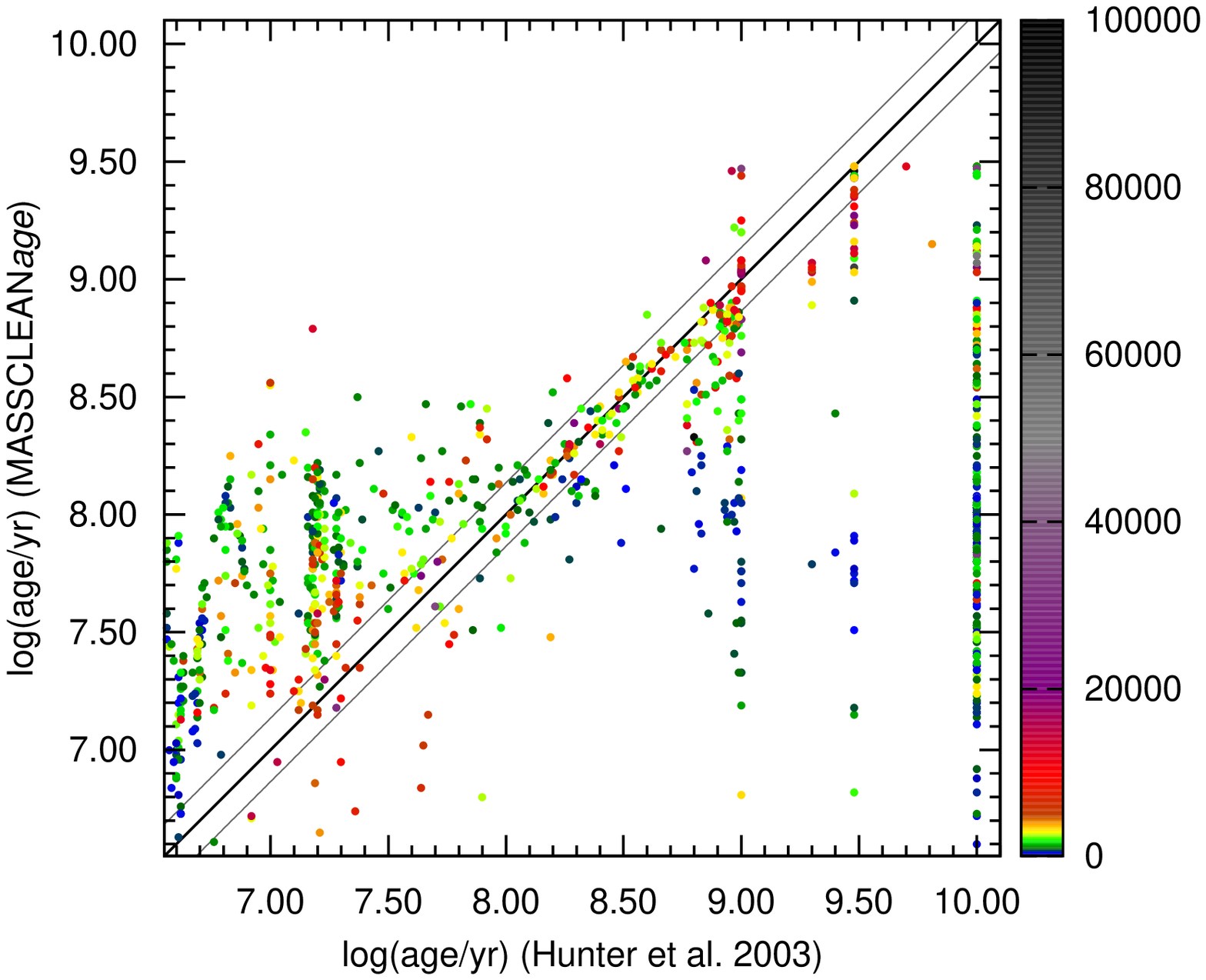}
\vspace{-0.7 cm}

\hspace{0.5 cm}
\includegraphics[angle=270,width=0.353\textwidth, bb= 280 144 330 680]{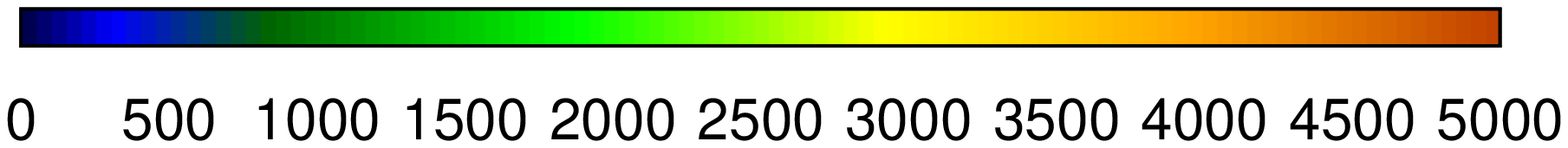}
\caption[]{\footnotesize Comparison of \citeauthor*{hunter2003} \citeyear{hunter2003} ages with MASSCLEAN ages.  {\it Top:} The dots are color-coded to display the dispersion in $(U-B)_{0}$ with respect to the {\it infinite mass limit}. Black dots represent clusters who lie very close to the colors from traditional SSP models computed in the {\it infinite mass limit}.  {\it Bottom:} The dots are color-coded to show the mass of the cluster as derived by MASSCLEAN{\fontfamily{ptm}\selectfont \textit{age}}.  \normalsize}\label{fig:paper4-06new}
\end{figure}

Secondly, in the lower panel of Figure \ref{fig:paper4-06new}, we see that most of the clusters lying away from the identity line are very low mass clusters (blue or green in color).  These small clusters can at times show very red colors (if a luminous red giant or supergiant happens to exist) mimicking a much older cluster of higher mass.  They also can at times show very blue colors (if it is caught without a single luminous, evolved red star) mimicking a younger cluster.  If one concentrates on the more massive clusters, the flattened slope is greatly reduced and the clusters are more equally scattered to either side of the identity line.  With the exception of some non-physical, vertical structures in the figure due to quantized age assignments by \citeauthor*{hunter2003} \citeyear{hunter2003}, the distribution of ages between the two methods are reasonably smooth over age, without any severe gaps or over-densities over the range from $log(age/yr)$ $=$ $6.5$ to $9.0$.  In particular, the upper panel of Figure \ref{fig:paper4-06new} shows that clusters which lie close to the  {\it infinite mass limit} (in black), are closely aligned with the ages given by MASSCLEAN{\fontfamily{ptm}\selectfont \textit{age}} over the range $log(age/yr) = 8.0 - 9.0$.

\begin{figure}[htp]
\centering
\includegraphics[angle=270,width=0.5\textwidth, bb= 115 170 525 650]{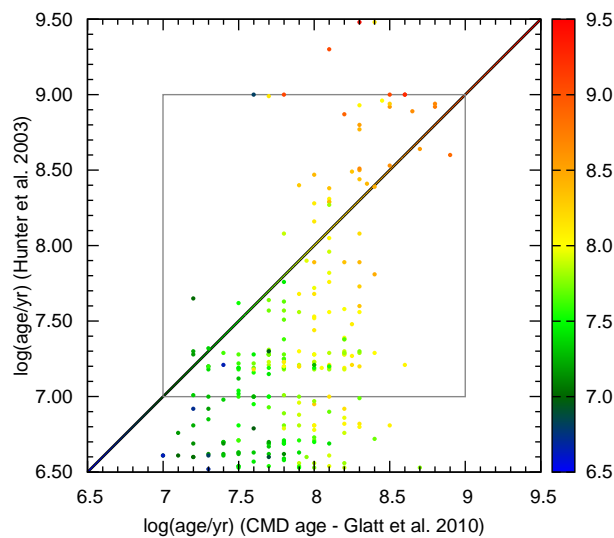}
\vspace{-0.3 cm}

\caption[]{\footnotesize Comparison of \citeauthor*{hunter2003} \citeyear{hunter2003} ages with CMD ages from \citeauthor*{glatt} \citeyear{glatt}.  The dots are color-coded to display the MASSCLEAN ages, and the same color coding is used for the identity line. The gray box indicates the age range \citeauthor*{glatt} \citeyear{glatt} estimated for this sample, similarly to Figure \ref{fig:paper4-04new}. \normalsize}\label{fig:paper4-07new}
\end{figure}

How do the \citeauthor*{hunter2003} \citeyear{hunter2003} ages compare to the \citeauthor*{glatt} \citeyear{glatt} CMD ages?  This is given in Figure \ref{fig:paper4-07new}. Consistent with the comparison to MASSCLEAN in Figure \ref{fig:paper4-06new}, quite a few clusters are found by \citeauthor*{hunter2003} \citeyear{hunter2003} to have very young ages, while the CMD ages for those clusters are as much as 10 or more times older.  As we explained previously, this can be understood in part by the use of Geneva models by \citeauthor*{hunter2003} \citeyear{hunter2003} compared to the Padova models used by the \citeauthor*{glatt} \citeyear{glatt} study.

\subsection{Age Determination Using a Traditional $\chi^2$ Minimization Method}

Perhaps the most common method for cluster age determinations from integrated colors uses a $\chi^{2}$ minimization method.  This method searches for the best match between a cluster's integrated colors to colors predicted by SSP models in the {\it infinite mass limit}.  This has been used to derive stellar cluster ages in the Small and Large Magellanic Clouds (\citeauthor*{rz05} \citeyear{rz05}; \citeauthor*{deGrijs2006} \citeyear{deGrijs2006}; \citeauthor*{chandar2010a} \citeyear{chandar2010a}), The Milky Way (\citeauthor*{hancock} \citeyear{hancock}), and most extensively to age star clusters found in external galaxies (\citeauthor*{deGrijs2003} \citeyear{deGrijs2003}; \citeauthor*{bik2003} \citeyear{bik2003}; \citeauthor*{bast05} \citeyear{bast05}; \citeauthor*{fall2005} \citeyear{fall2005}; \citeauthor*{Kaleida2010} \citeyear{Kaleida2010}; \citeauthor*{bast12} \citeyear{bast12}).  We will now investigate this method based on the \citeauthor*{hunter2003} \citeyear{hunter2003} $UBVR$ stellar cluster photometry in the LMC and compare the ages with the \citeauthor*{glatt} \citeyear{glatt} CMD ages. 

The question of how to deal with the unknown extinction arises when deriving cluster age with the $\chi^2$ method.  For many studies, this has been included within the minimization fit (\citeauthor*{bast05} \citeyear{bast05}; \citeauthor*{fall2005} \citeyear{fall2005}; \citeauthor*{hancock} \citeyear{hancock}; \citeauthor*{chandar2010a} \citeyear{chandar2010a}).  However, some authors choose to estimate extinction first independently, before running the minimization method to derive cluster age (de Grijs \& Anders 2006, Rafelski \& Zaritsky 2005).  In particular, \citeauthor*{rz05} \citeyear{rz05} found including the extinction as a free parameter in their fitting routine lead to unreliable estimates of extinction.  We agree that extinction is best handled separately, as there is no causal relationship between extinction and age.  We apply the extinction available for each cluster from \citeauthor*{glatt} \citeyear{glatt} or assume a global extinction of $E(B-V) = 0.13$, as used by \citeauthor*{hunter2003} \citeyear{hunter2003}.  We then derive the best-fit values of age that minimize the difference between these de-reddened observed colors and the model colors, respectively, and run this simultaneously over the bands, $U, B, V, R$.  The routine finds the location along the traditional SSP {\it infinite mass limit} line where the stellar cluster is seen to most closely line up in the three color bands, $(U-B)_{0}$, $(B-V)_{0}$, and $(V-R)_{0}$.  In such a fit, integrated $M_{V}$ magnitudes are not used in deriving age because traditional models provide for only mass-insensitive colors as a function age.  

\begin{figure}[htp]
\centering
\includegraphics[angle=270,width=0.5\textwidth, bb= 115 170 525 650]{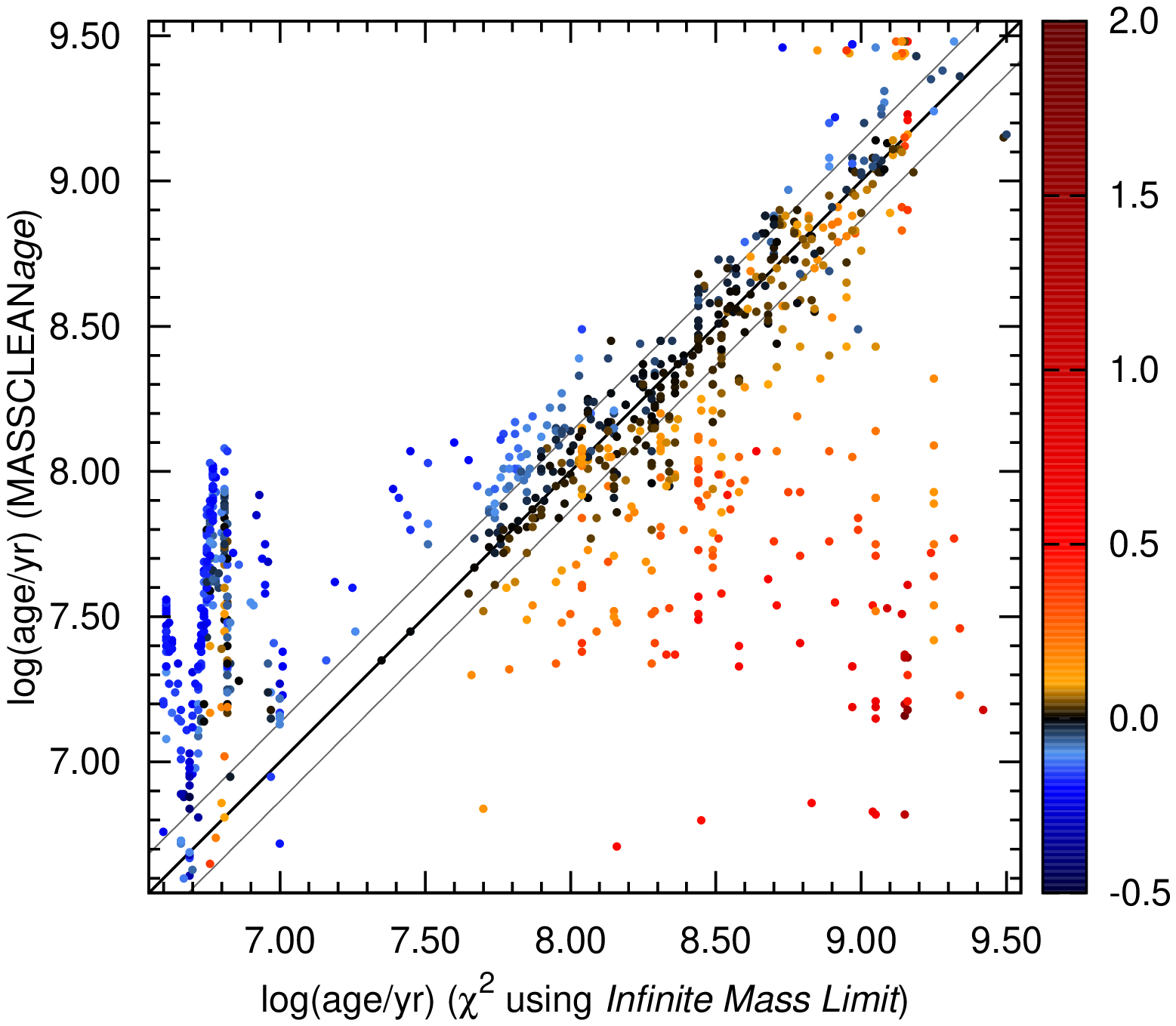}

\includegraphics[angle=270,width=0.5\textwidth, bb= 115 170 525 650]{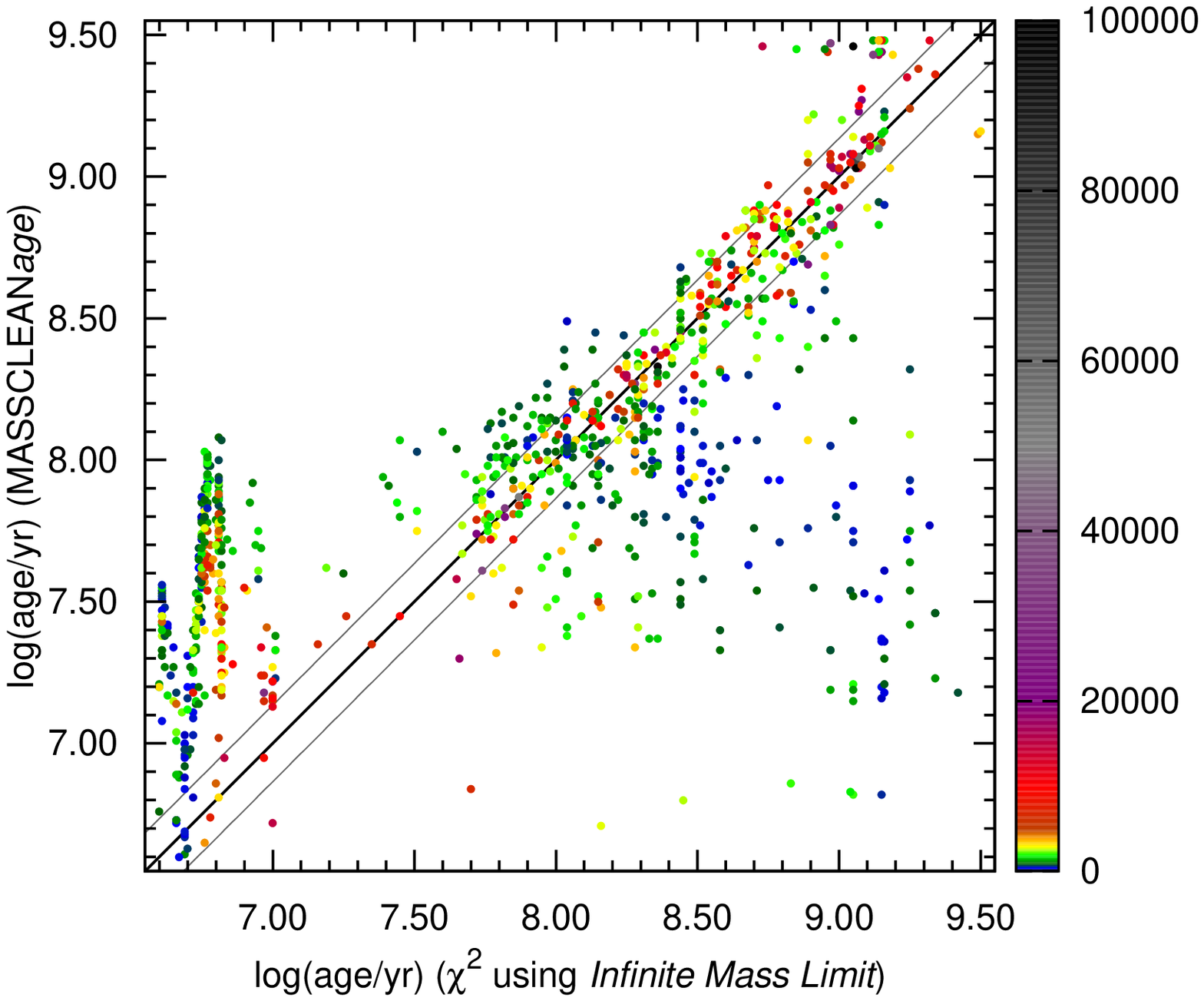}
\vspace{-0.7 cm}

\hspace{0.5 cm}
\includegraphics[angle=270,width=0.353\textwidth, bb= 280 144 330 680]{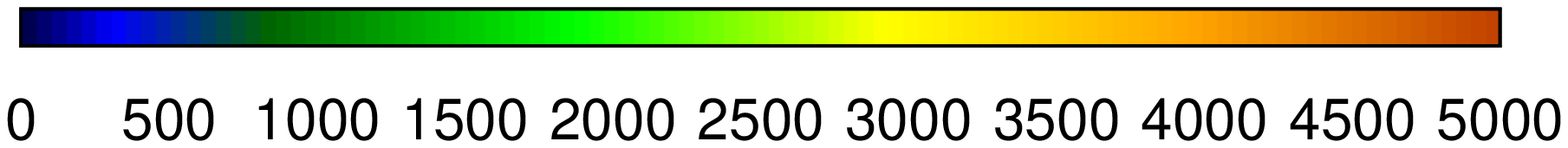}
\caption[]{\footnotesize MASSCLEAN{\fontfamily{ptm}\selectfont \textit{age}} results versus a $\chi^2$ minimization fit to traditional SSP models assuming an {\it infinite mass limit}. {\it Top:} The dots are color-coded to display the dispersion in $(U-B)_{0}$ with respect to the {\it infinite mass limit}. {\it Bottom:} The dots are color-coded to show the mass of the cluster.  \normalsize}\label{fig:paper4-08new}
\end{figure}

\begin{figure}[htp]
\centering
\includegraphics[angle=270,width=0.5\textwidth, bb= 115 170 525 650]{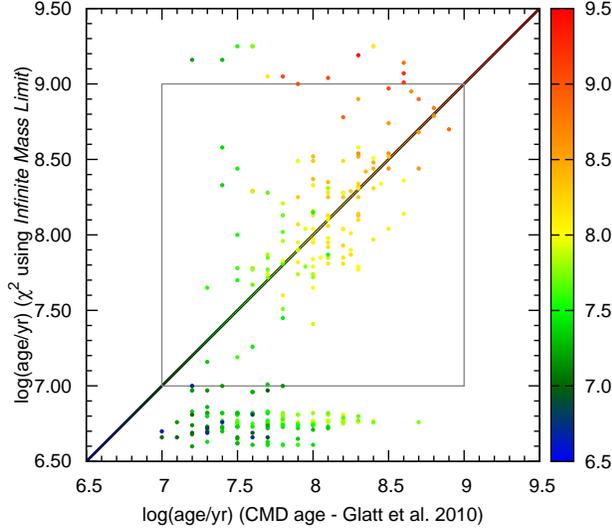}
\vspace{-0.3 cm}

\caption[]{\footnotesize Comparison of ages determined by traditional $\chi^2$ minimization using the {\it infinite mass limit} with CMD ages from \citeauthor*{glatt} \citeyear{glatt}. The dots are color-coded to display the MASSCLEAN ages.  \normalsize}\label{fig:paper4-09new}
\end{figure}

How do the $\chi^2$ age determination methods compare to ages determined using MASSCLEAN{\fontfamily{ptm}\selectfont \textit{age}}?   In Figure \ref{fig:paper4-08new}, we present the comparison between the MASSCLEAN{\fontfamily{ptm}\selectfont \textit{age}} results and the ages computed using the $\chi^{2}$ minimization method in the {\it infinite mass limit}.   In the top panel the colors of the dots show the dispersion in $(U-B)_{0}$ with respect to the {\it infinite mass limit} such as shown in the top panel of Figure \ref{fig:paper4-06new}. In the bottom panel the colors display the mass of the cluster.  Here we see that the vast majority of clusters lying away from the identity line are very low mass clusters (blue or green in color).  We see a similar steepening in the slope as seen with the \citeauthor*{hunter2003} \citeyear{hunter2003} comparison: most of the very young clusters as defined by $\chi^2$ minimization, are given older ages using MASSCLEAN{\fontfamily{ptm}\selectfont \textit{age}}.  

How do the $\chi^2$ ages compare to the \citeauthor*{glatt} \citeyear{glatt} CMD ages?  This is given in Figure \ref{fig:paper4-09new}. The dots are color-coded to show the MASSCLEAN ages, as well as the identity line. Consistent with the comparison to MASSCLEAN, in Figure \ref{fig:paper4-09new} quite a few clusters are found to have very young ages by the $\chi^2$ method, while the CMD ages for those clusters are as much as 10 or more times older.  A large density of clusters is found in the $log(age/yr)$ $=$ $6.5$ to $7.0$, followed by a dearth of clusters in the range $log(age/yr)$ from $7.0$ to $7.5$ using the $\chi^{2}$ method.  A similar structure hinting at the same problem is seen comparing $\chi^2$ ages with other age determinations obtained in M51 in \citeauthor*{bast11} \citeyear{bast11} (see their Figure 1).  Such an age distribution is not supported by the \citeauthor*{glatt} \citeyear{glatt} CMD results, which show a smooth distribution with age over these times.   This effect, seen very clearly in the Figures \ref{fig:paper4-08new} and \ref{fig:paper4-09new}, shows up in the literature when researchers use a $\chi^{2}$ minimization method (e.g. \citeauthor*{bast05} \citeyear{bast05}; \citeauthor*{fall2005} \citeyear{fall2005}; \citeauthor*{deGrijs2006} \citeyear{deGrijs2006}; \citeauthor*{chandar2010a} \citeyear{chandar2010a}).  \citeauthor*{bik2003} \citeyear{bik2003} and \citeauthor*{Gieles2005} \citeyear{Gieles2005}, using a similar $\chi^2$ method to derive the cluster age distribution for M51 stellar clusters, looked closely at their age distribution and acknowledged that density features such as these coming from their analysis.  They attributed it to properties of the cluster models and the age determination methods they employed ($\chi^2$ minimization).  We note, both of these studies are limited to fairly massive clusters in this distant galaxy, M51, indicating {\bf high mass clusters are not immune to this effect} when using a $\chi^2$ method.   As it is revealed here in Figures \ref{fig:paper4-08new} and \ref{fig:paper4-09new}, this effect will greatly influence subsequent analyses based on these incorrect age distributions.  

\subsection{Why the Dearth of Clusters from $log(age/yr) = 7.0$ to $7.5$ Using $\chi^2$ Minimization?}

What is happening to cause such a defective age distribution from the $\chi^2$ analysis?  The problem may in part lie with some of the assumptions made in applying the method.  For instance, many groups are assuming a Gaussian distribution of values about the mean (\citeauthor*{Dolphin2002} \citeyear{Dolphin2002}), which is required for a $\chi^{2}$ minimization method to be valid.  However, moderate and low mass stellar clusters (mass at or below $10^{4}$ $M_{\Sun}$) do not exhibit a Gaussian distribution about the model expectation value (e.g. \citeauthor*{lancon2000} \citeyear{lancon2000}, \citeyear{lancon2002}; \citeauthor*{paper2} \citeyear{paper2}, \citeyear{paper3};  \citeauthor*{lancon2010} \citeyear{lancon2010}).  There is also a large scatter of clusters older than $log(age/yr) > 8.00$, as defined by the $\chi^2$ minimization method, which MASSCLEAN{\fontfamily{ptm}\selectfont \textit{age}} has determined to be $log(age) < 8.00$.  Again, many of the clusters that lie far from the identity line are low mass, as seen in the lower panel of Figure \ref{fig:paper4-08new}.  However even relatively high mass clusters (yellow and orange) are found to lie far from the identity line.  The answer lies in Figure \ref{fig:paper4-11new}  (a) and (b).

In Figures \ref{fig:paper4-11new}  (a) and (b) we have plotted our MASSCLEAN{\fontfamily{ptm}\selectfont \textit{age}}-derived ages and the $\chi^2$-derived ages, respectively, as a function of observed $(U-B)_{0}$ color.  The colors of the dots in Figure \ref{fig:paper4-11new} (b) show the corresponding age computed by MASSCLEAN{\fontfamily{ptm}\selectfont \textit{age}}, while the dots are placed on the figure according to their $\chi^2$-derived ages. The {\it infinite mass limit} (continuous line) is also colored to show the MASSCLEAN-derived age. As a background in Figure \ref{fig:paper4-11new} we also show in gray the range in integrated $(U-B)_{0}$ colors that were reproduced with our 100 million Monte Carlo simulations, for clusters in the $200-100,000$ $M_{\Sun}$ range.  This demonstrates the colors one can expect to find moderate mass stellar clusters to lie within this diagram.

\begin{figure}[htp]
\centering
\subfigure[]{\includegraphics[angle=270,width=0.5\textwidth, bb= 115 75 525 740]{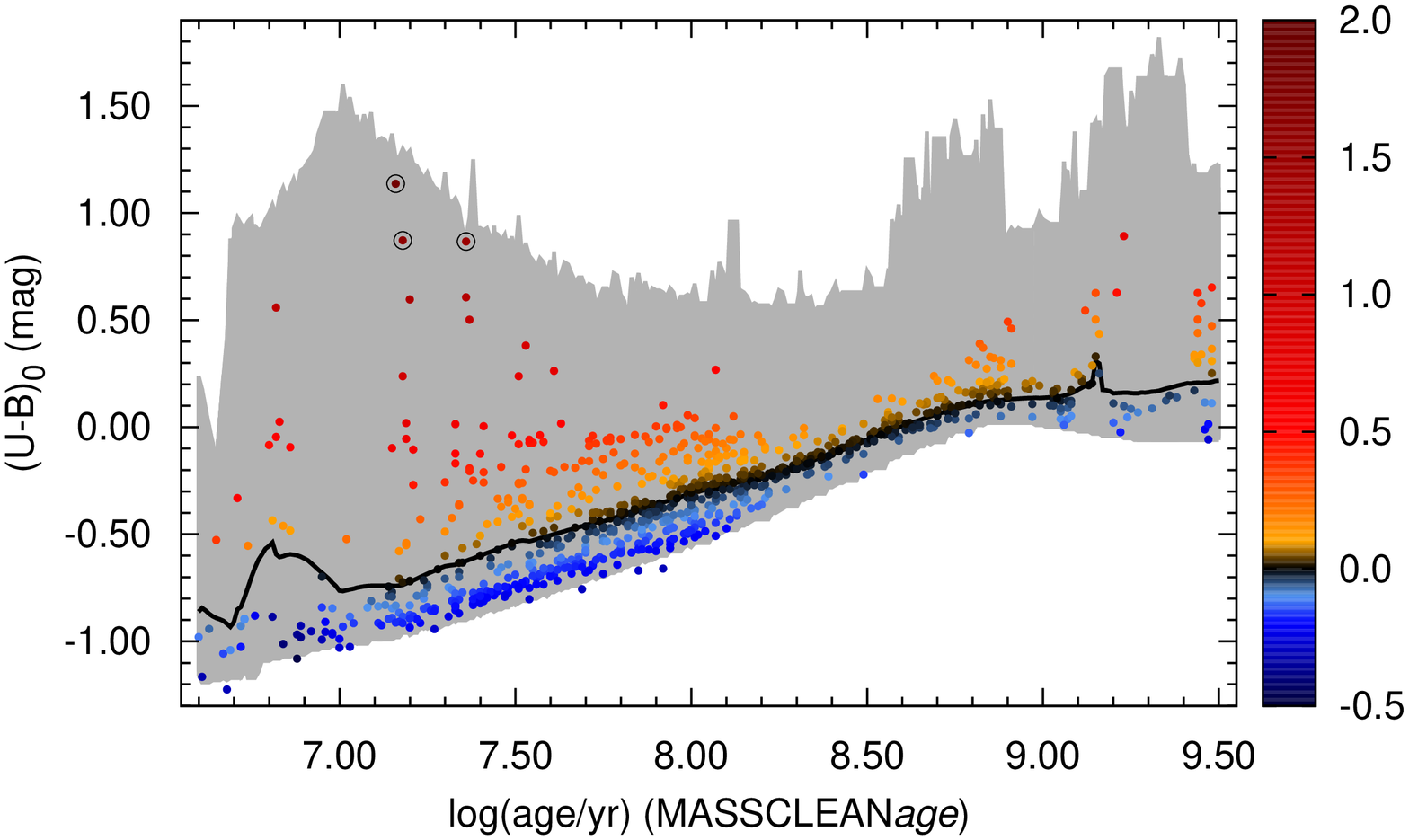}} 
\subfigure[]{\includegraphics[angle=270,width=0.5\textwidth, bb= 115 75 525 740]{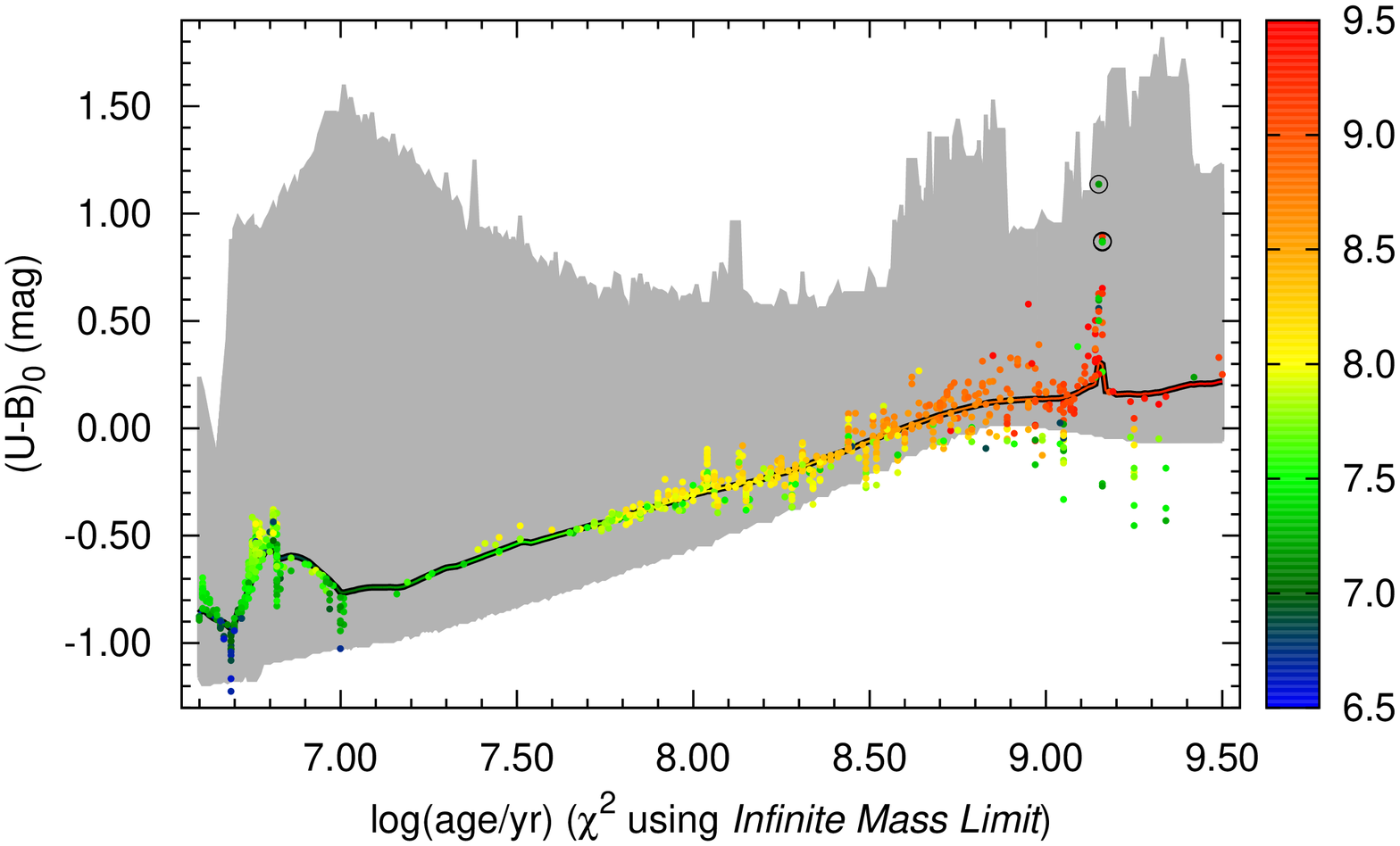}}
\caption[]{\footnotesize $(U-B)_{0}$ vs. $log(age/yr)$ for 920 LMC clusters. The dispersion range from the MASSCLEAN{\fontfamily{ptm}\selectfont \textit{colors}} database is presented is gray. (a) The ages are computed using MASSCLEAN{\fontfamily{ptm}\selectfont \textit{age}} and the dots are color-coded to show the level of dispersion in $(U-B)_{0}$ with respect to the {\it infinite mass limit} (black line). (b) The ages are computed using the traditional method of $\chi^{2}$ minimization, based on the {\it infinite mass limit}. The dots are color-coded to display the corresponding age computed by MASSCLEAN{\fontfamily{ptm}\selectfont \textit{age}}. Three extremely red clusters are highlighted by black circles illustrating the difference between the age derived by the two methods. (Note that the lower circle in (b) contains two clusters).  \normalsize}\label{fig:paper4-11new}
\end{figure}

In the Figure \ref{fig:paper4-11new} (a) the colors of the dots are chosen to display the dispersion in $(U-B)_{0}$ with respect to the {\it infinite mass limit} (black line).  For ages derived using MASSCLEAN{\fontfamily{ptm}\selectfont \textit{age}}, the $(U-B)_{0}$ color can vary greatly from the {\it infinite mass limit} model line.   Conversely, because the $\chi^2$ method seeks to place clusters {\sl on} the line (or close to it), we see a very small dispersion of clusters from the predicted {\it infinite mass limit} model line in Figure \ref{fig:paper4-11new} (b).   The range of colors observed in stellar clusters is not due to photometric error, but due to stochastic variations populating the upper end of the stellar mass function and leading to non-Gaussian color ranges.   Given what we see in Figure \ref{fig:paper4-10new} and knowing the clusters will not always lie on the line, how can forcing the clusters on to the line be the right method?   

Looking at Figure \ref{fig:paper4-11new} (a) and (b), we can clearly see why a dearth in clusters appears in the age range $log(age/yr)$ from $7.0$ to $7.5$ using the $\chi^{2}$ method.  Clusters in the Figure \ref{fig:paper4-11new} (a) lying below the traditional SSP line that is computed in the {\it infinite mass limit}, have no evolved stars and are very blue for their age.  They will all be matched to young ages in $\chi^{2}$ modeling, so they will lie upon the traditional SSP line.  This causes both a pile up of younger clusters, as well as a loss of clusters between $7.0$ and $7.5$ due to the 'turn up' of $(U-B)_{0}$ color at  $log(age/yr) = 7.0$ (the red giant bump).  Likewise, clusters that naturally lie above the traditional SSP line, when computed with the $\chi^{2}$ method in the {\it infinite mass limit}, are forced to the right, and made older, or to the left, if they are not too red to be placed on the 'bump' centered at $log(age/yr) = 6.8$.  Both effects will preferentially empty out the age range  $log(age/yr)$ $7.0$ to $7.5$.   The traditional $\chi^2$ method gives both too old and too young of ages for clusters that should naturally fall in the range between $log(age/yr)$ from $7.0$ to $7.5$ simply because of the red giant bump, in the shape of the {\it infinite mass limit} line.

Another feature in Figure \ref{fig:paper4-11new} (b) gives us cause for concern.  There are a number of old clusters that have very blue colors.  They lie outside the 'observable' gray zone established by MASSCLEAN.  The lower blue limit in Figure \ref{fig:paper4-11new} (a) and (b) are hard limits representing the bluest color possible where virtually all stars are on the main sequence for the corresponding cluster age with mass above $200$ $M_{\Sun}$. Clusters bellow $200$ $M_{\Sun}$ will have a slightly lower limit, but they are below the fading limit (see Section \ref{fading}) for such ages, so they would not be detected.  Anything bluer (outside the photometric errors limit) is simply non-physical.  Note, the clusters found in the non-physical region are mostly made up of the missing $log(age/yr)$ $7.0$ to $7.5$ clusters. The traditional $\chi^{2}$ minimization method, due to the variation of $U,B,V,R$ colors, will always produce an overdensity of younger and older clusters, with an underdensity of clusters in the $log(age/yr)=[7.0,7.5]$ range.

\subsection{Age Determination for Stellar Clusters Exhibiting Extreme Colors}

 In Figure \ref{fig:paper4-10new} we show the full sample of 920 LMC clusters in the observed $UBV$ color-color plane, with the standard SSP, {\it infinite mass limit} model line running diagonally through the middle. In here in lies the challenge with age derivations using integrated colors. Clusters often lie a significant distance from this line.  While we can only show two colors here, this is true no matter how many colors one tries to measure. How does MASSCLEAN{\fontfamily{ptm}\selectfont \textit{age}} perform when asked to determine the age of clusters with the most extreme colors, far from the standard, {\it infinite mass limit}, SSP color expectation? 

In Figures \ref{fig:paper4-11new} (a) and (b) three clusters are highlighted by black circles (In Figure \ref{fig:paper4-11new} (b) only two circles are clearly visible, the lower one contains two clusters).  These clusters were selected for the sole reason they are the most red clusters lying furthest from the expected {\it infinite mass limit} model predictions in $(U-B)_{0}$.  They make an interesting first test of how the two methods handle real, but extreme, observed colors.  The $\chi^{2}$ minimization predicts all of them as being quite old clusters of moderately high mass, $17,000-32,000$ $M_{\Sun}$.   However, MASSCLEAN{\fontfamily{ptm}\selectfont \textit{age}} finds a very different solution.  The logic here is that such extreme colors can only come from a very low mass system, where stochastic fluctuations dominate how the stellar mass function is populated.  The color dispersion is instead matched with a cluster mass of only $200-300$ $M_{\Sun}$ with MASSCLEAN{\fontfamily{ptm}\selectfont \textit{age}}.  Thus a very young age, 15-25 million years is found versus the 1.4 billion years found via the $\chi^{2}$ minimization method.   Here we see there is more than just a drastic disagreement between the methods in deriving the ages of these clusters.  There exists an equally drastic disagreement about the mass of these clusters, a result that is as disturbing as the very large uncertainty in age.

\begin{figure}[htp]
\centering
\includegraphics[angle=0,width=0.495\textwidth, bb=  65 120 557 670]{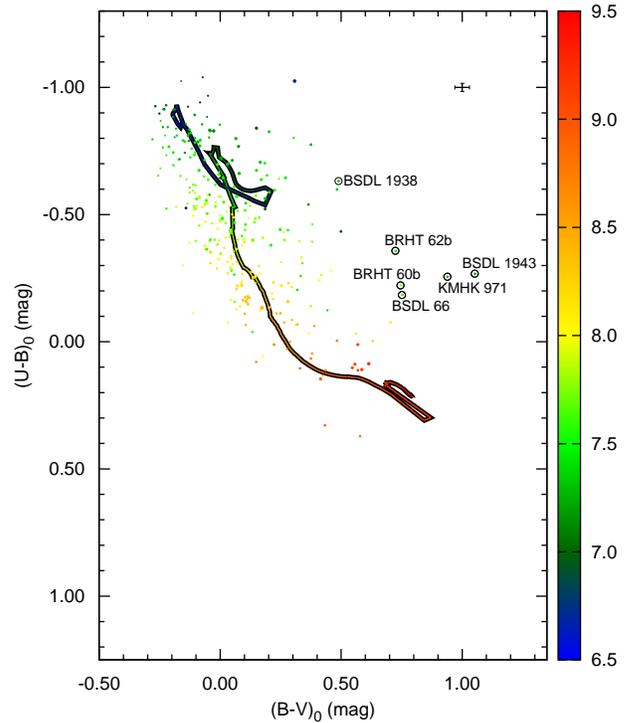}

\vspace{-0.4 cm}

\caption{\footnotesize  LMC clusters with CMD age on $(U-B)_{0}$ vs. $(B-V)_{0}$ color-color diagram. The color is scaled with the age, and the size of the dots is scaled with the mass. The {\it infinite mass limit} also displays the age. The six clusters presented in Table \ref{table3} are highlighted by black circles and labels. The mean error is given in the upper right corner.  \normalsize}\label{fig:paper4-14new}
\end{figure}

In addition to the previous three examples, we now select six very red clusters (in $(B-V)_{0}$ and/or $(V-R)_{0}$) that have \citeauthor*{glatt} \citeyear{glatt} CMD ages.  These are presented in the Table \ref{table3}.  Their location, far away from the traditional {\it infinite mass limit}, is presented in the color-color diagram of Figure \ref{fig:paper4-14new} (\citeauthor*{hunter2003} \citeyear{hunter2003}).  The sample of extreme clusters are circled in black.  The remaining clusters shown in the diagram represent the entire subsample of 288 \citeauthor*{hunter2003} \citeyear{hunter2003} clusters that have \citeauthor*{glatt} \citeyear{glatt} CMD ages. The integrated colors shown here have been de-reddened using \citeauthor*{glatt} \citeyear{glatt} extinction. 

\begin{figure}[htp]
\centering
\includegraphics[angle=270,width=0.5\textwidth, bb= 100 100 520 770]{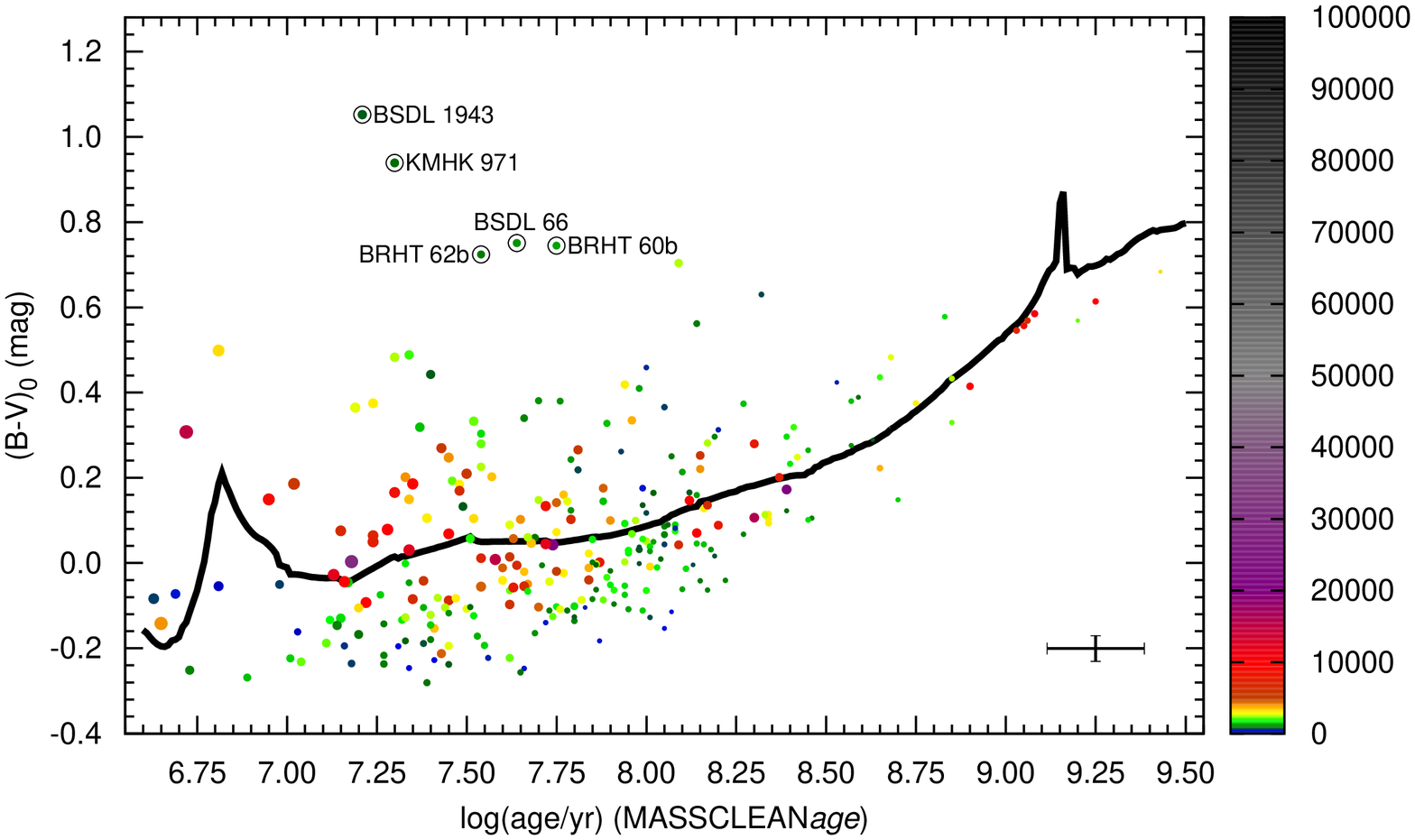}

\vspace{-0.15 cm}
\includegraphics[angle=270,width=0.395\textwidth, bb= 285 144 330 680]{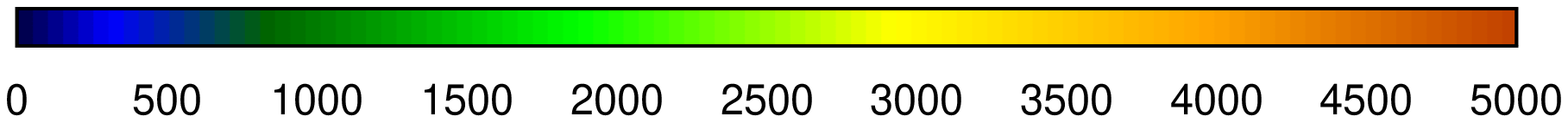}


\vspace{-0.4 cm}
\includegraphics[angle=270,width=0.5\textwidth, bb= 100 100 520 770]{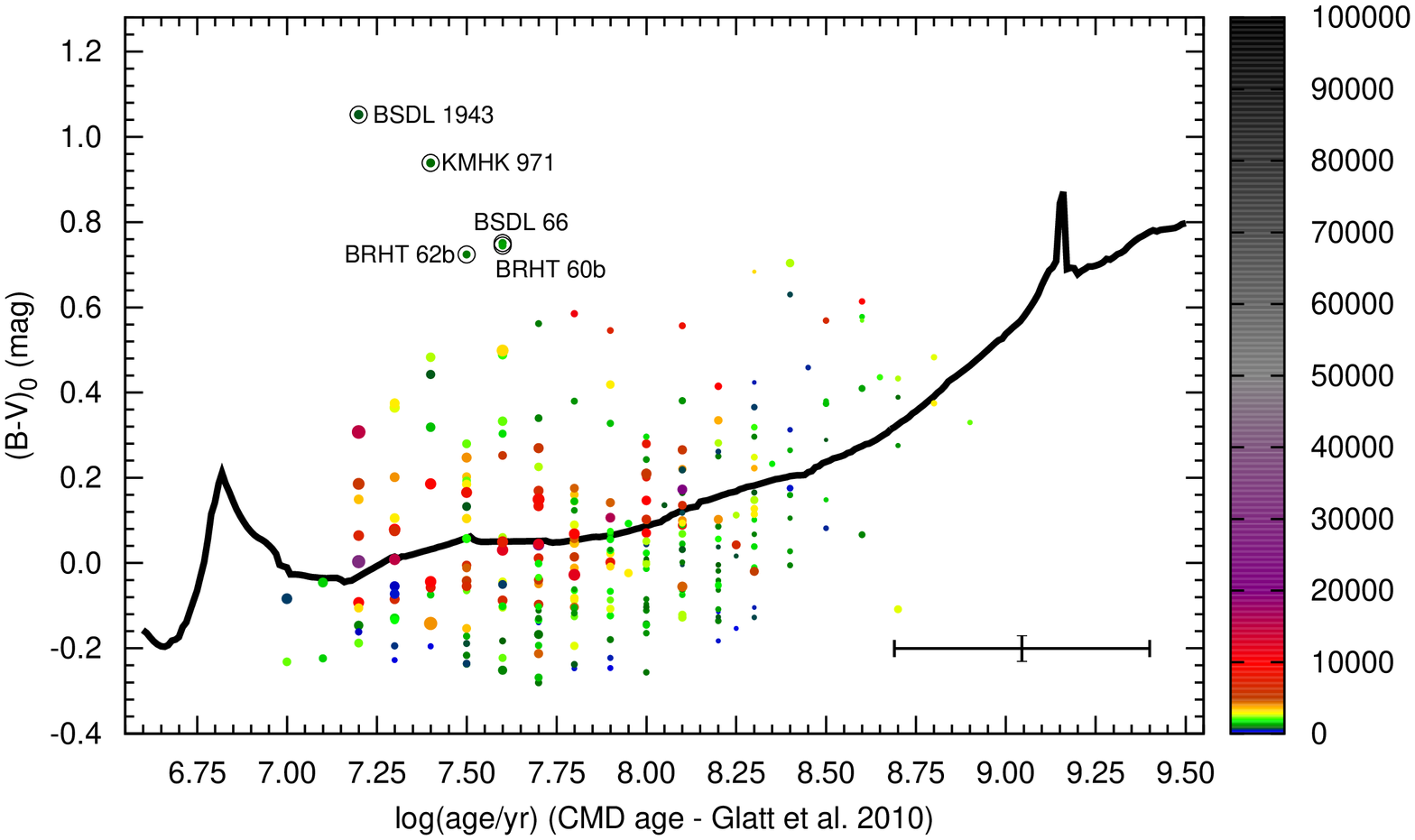}

\vspace{-0.15 cm}
\includegraphics[angle=270,width=0.395\textwidth, bb= 285 144 330 680]{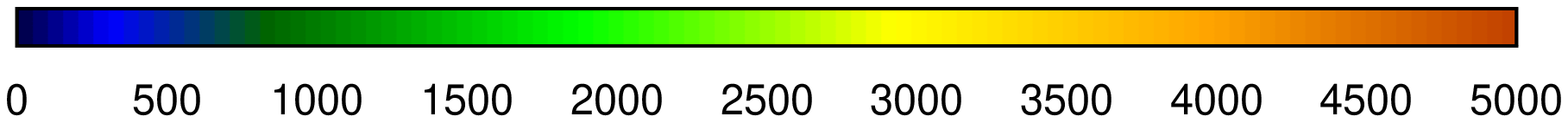}

\vspace{-0.4 cm}
\includegraphics[angle=270,width=0.5\textwidth, bb= 100 100 520 770]{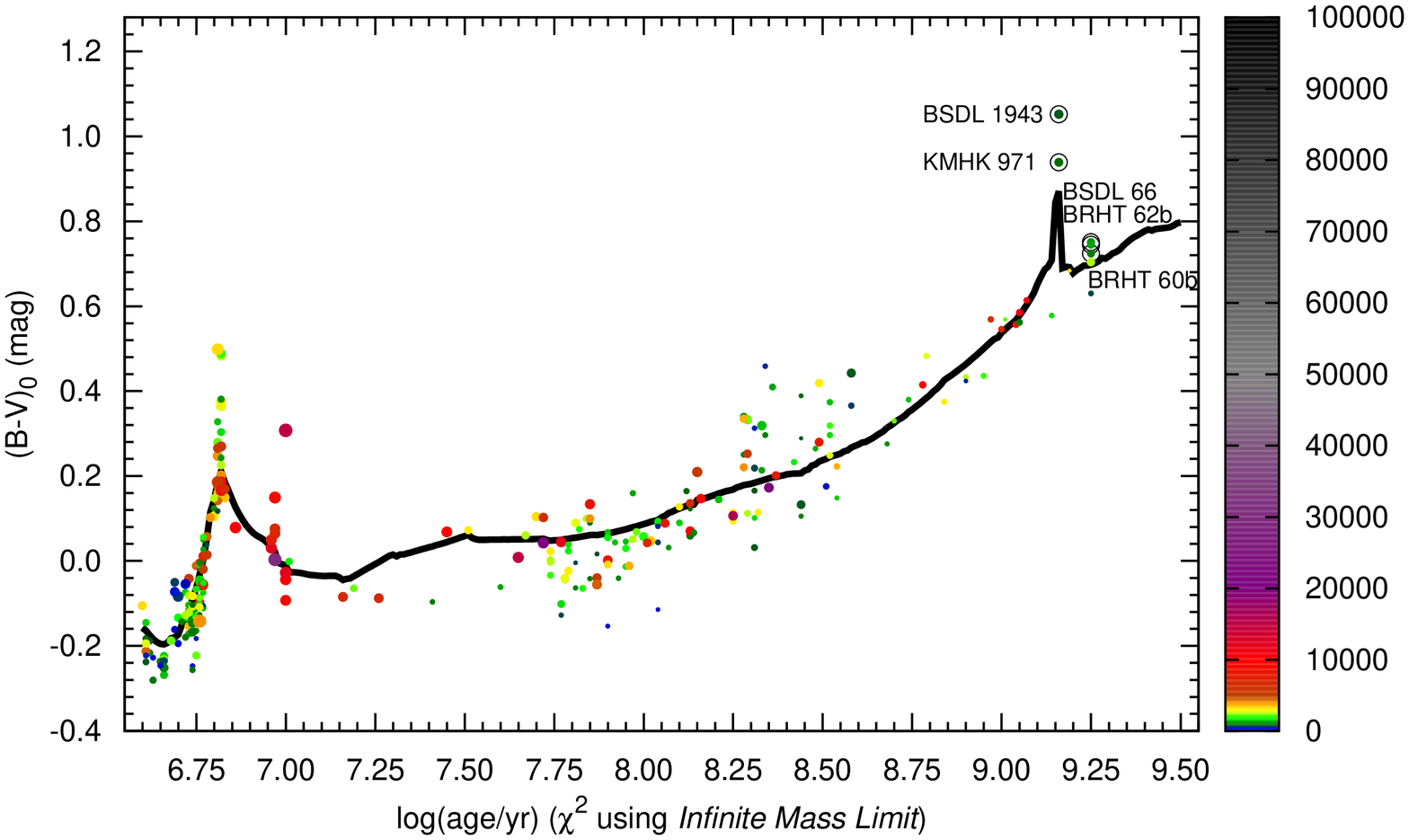}

\vspace{-0.15 cm}
\includegraphics[angle=270,width=0.395\textwidth, bb= 285 144 330 680]{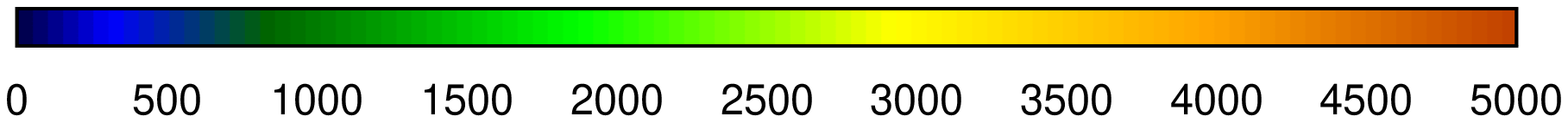}


\vspace{-0.3 cm}
\caption[]{\footnotesize Integrated $(B-V)_{0}$ colors vs. $log(age/yr)$ for 288 LMC clusters with CMD age. The {\it infinite mass limit} for Padova $Z=0.008$ is presented as a black line. The dots are color-coded based on the cluster mass. Five red clusters are high-lighted by black circles and labels to compare the age determined by different methods. {\it Top:} MASSCLEAN{\fontfamily{ptm}\selectfont \textit{age}}. {\it Middle:} CMD age from \citeauthor*{glatt} \citeyear{glatt}. {\it Bottom:} Traditional $\chi^2$ minimization, fitting to SSP models using the {\it infinite mass limit}. The mean error is given in the lower right corner. \normalsize}
\label{fig:paper4-12new}
\end{figure}

\begin{figure}[htp]
\centering
\includegraphics[angle=270,width=0.5\textwidth, bb= 100 100 520 770]{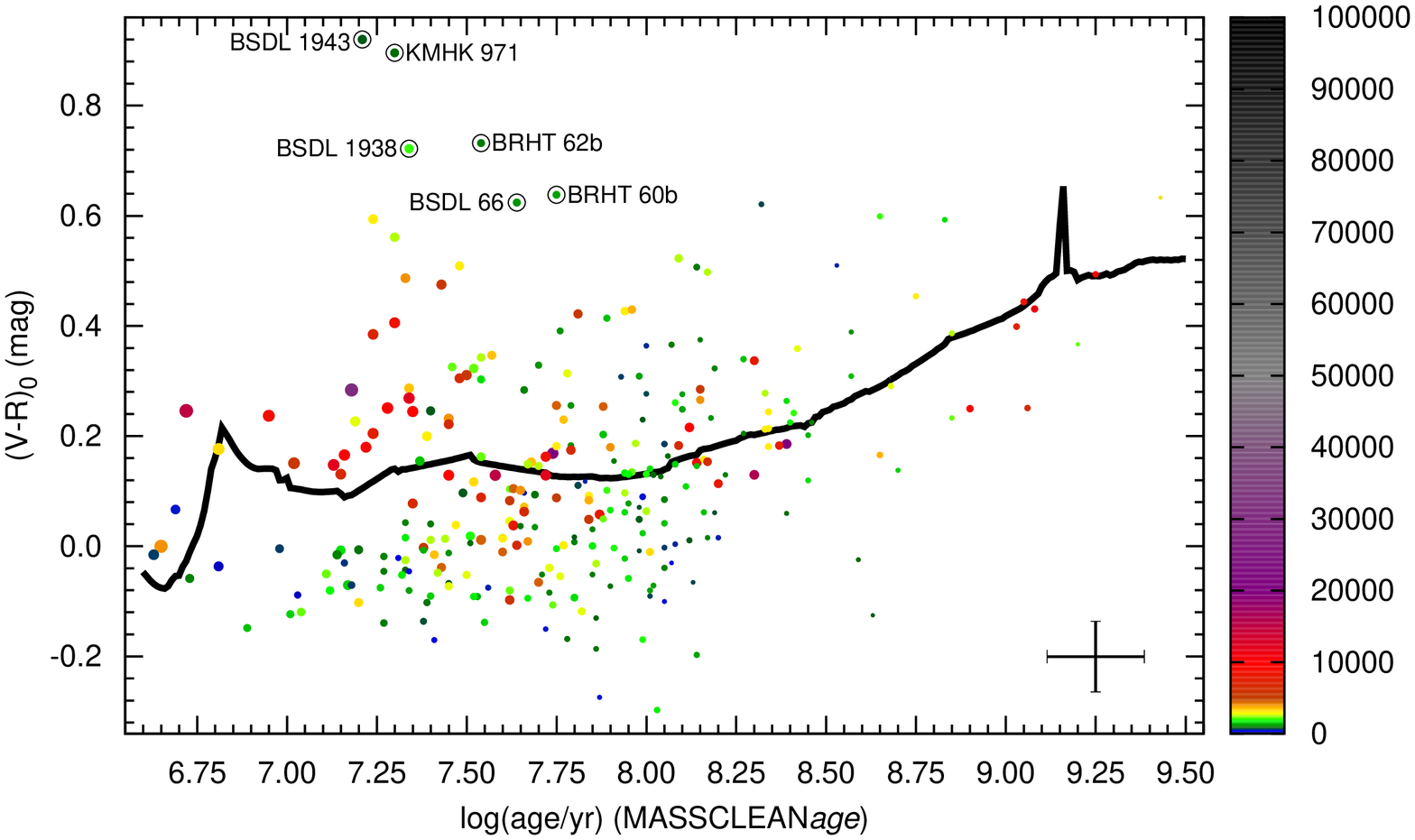}

\vspace{-0.15 cm}
\includegraphics[angle=270,width=0.395\textwidth, bb= 285 144 330 680]{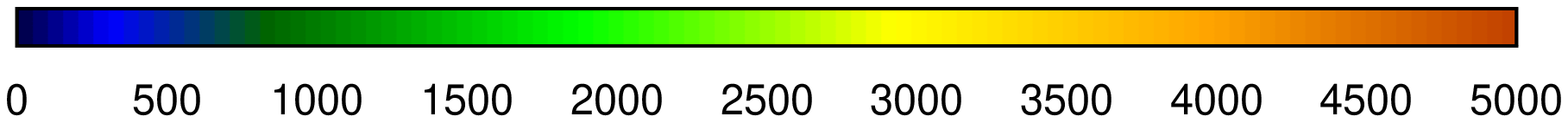}


\vspace{-0.4 cm}
\includegraphics[angle=270,width=0.5\textwidth, bb= 100 100 520 770]{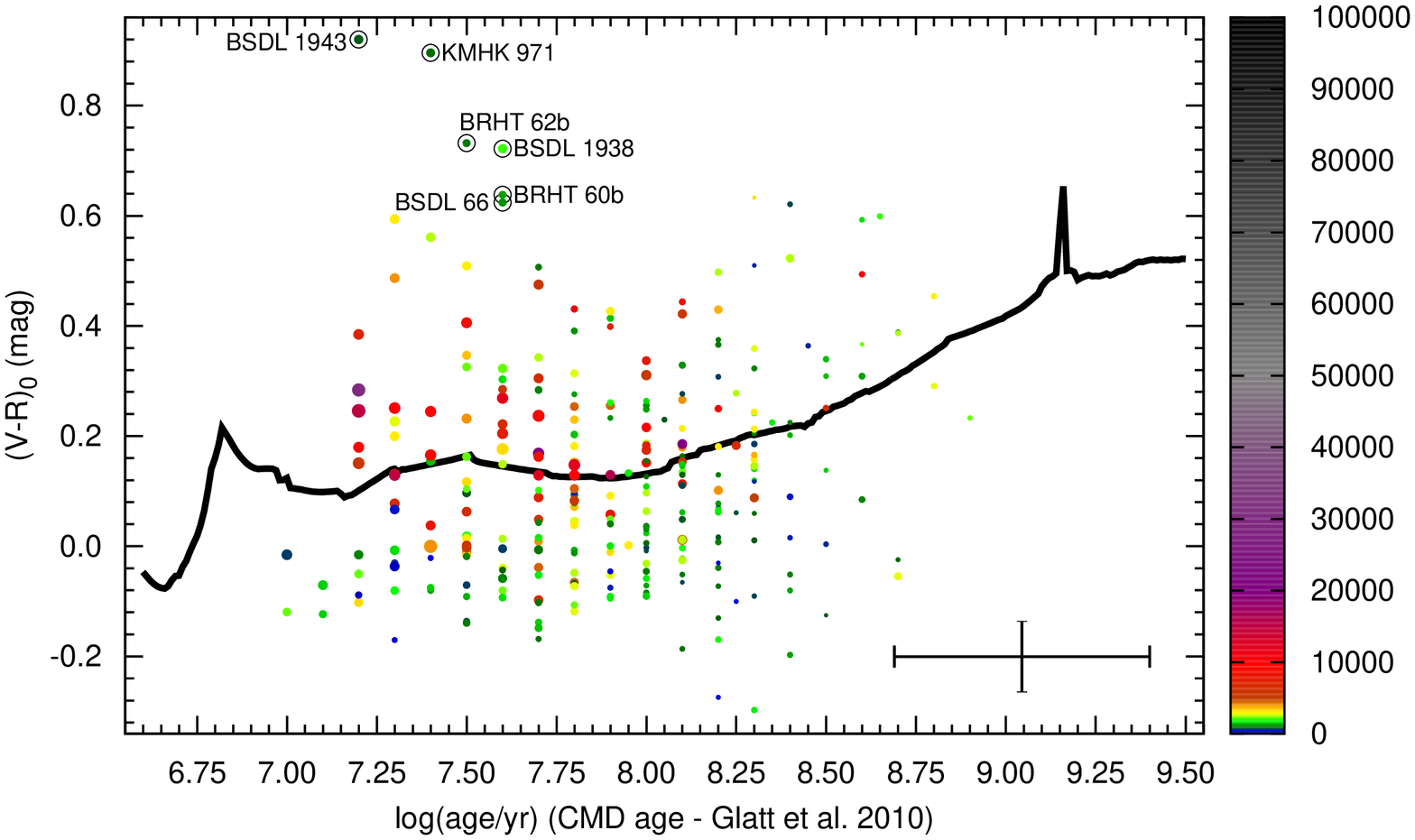}

\vspace{-0.15 cm}
\includegraphics[angle=270,width=0.395\textwidth, bb= 285 144 330 680]{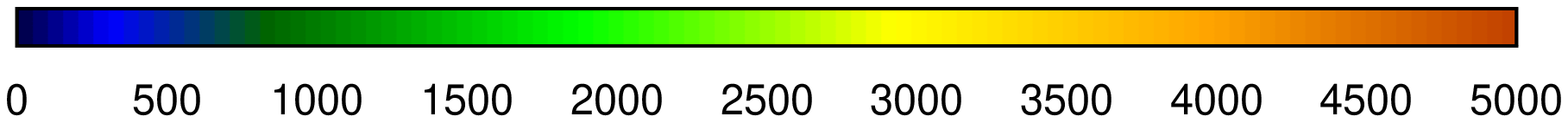}

\vspace{-0.4 cm}
\includegraphics[angle=270,width=0.5\textwidth, bb= 100 100 520 770]{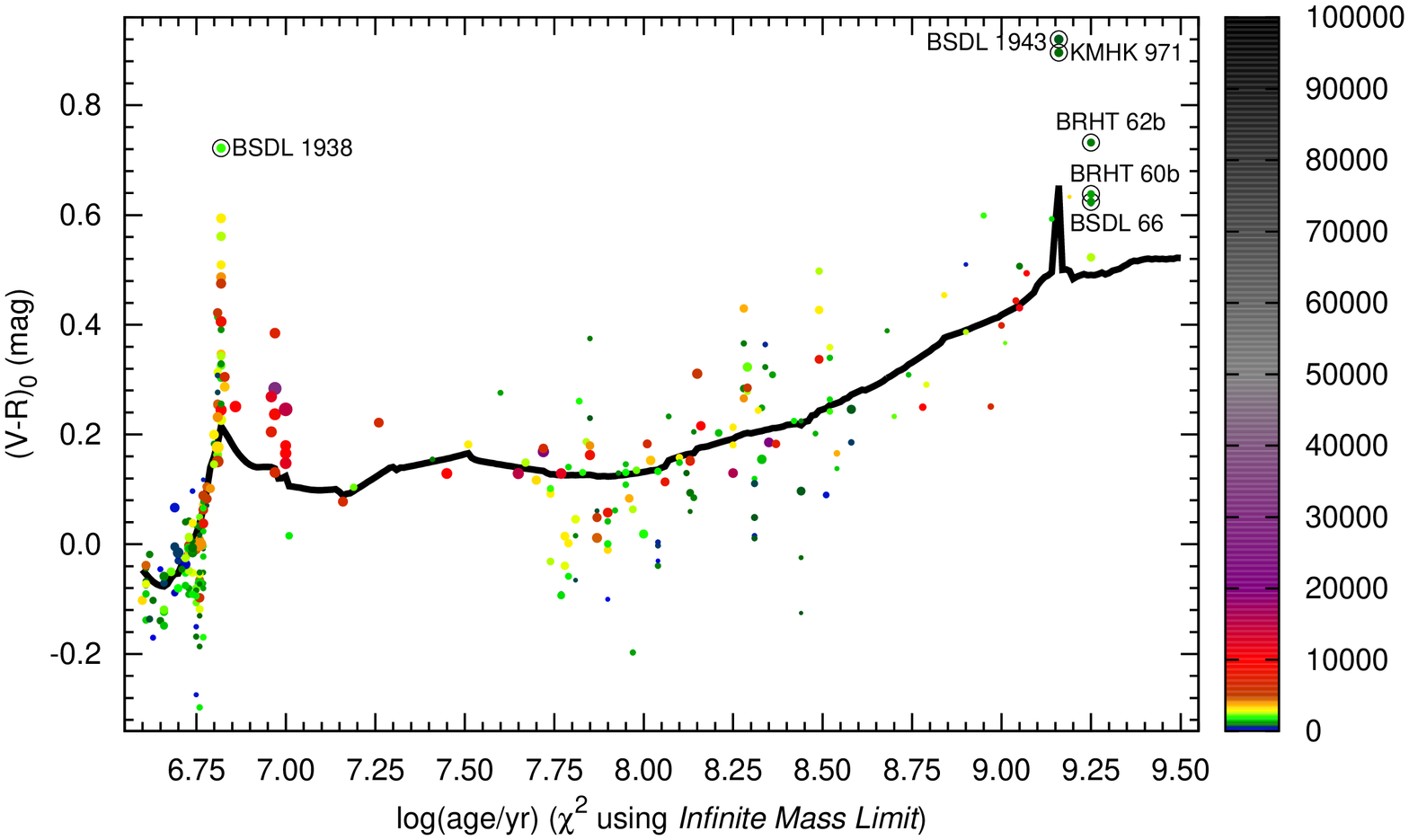}

\vspace{-0.15 cm}
\includegraphics[angle=270,width=0.395\textwidth, bb= 285 144 330 680]{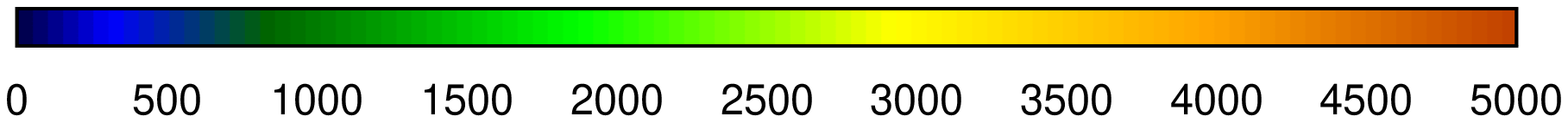}


\vspace{-0.3 cm}
\caption[]{\footnotesize Same as Figure \ref{fig:paper4-12new} but for $(V-R)_{0}$ vs. $log(age/yr)$. Six red clusters are highlighted by black circles and labels to compare the age determined by different methods. \normalsize}
\label{fig:paper4-13new}
\end{figure}

Five out of the six clusters, with the circled extreme clusters, $BDSL\phn 1943$, $KMHK\phn 971$, $BDSL\phn 66$, $BRHT\phn 60b$, and $BRHT\phn 62b$, are now plotted in Figures \ref{fig:paper4-12new} and \ref{fig:paper4-13new}. These five clusters are very red both in $(B-V)_{0}$ (Figure \ref{fig:paper4-12new}) and $(V-R)_{0}$ (Figure \ref{fig:paper4-13new}). In the top panels the ages computed by MASSCLEAN{\fontfamily{ptm}\selectfont \textit{age}} are presented. In the middle panels are presented the CMD ages from \citeauthor*{glatt} \citeyear{glatt}.  In the bottom panels are presented the ages derived using a traditional $\chi^{2}$ minimization with respect to the {\it infinite mass limit}. 

The MASSCLEAN{\fontfamily{ptm}\selectfont \textit{age}} extreme clusters agree fairly well with the \citeauthor*{glatt} \citeyear{glatt} ages; all are young with ages $log(age/yr) < 7.75$.  Moreover, the natural distribution of cluster locations in the diagram, showing that the clusters often lie far from the {\it infinite mass limit} line, is seen in both \citeauthor*{glatt} \citeyear{glatt} and the MASSCLEAN{\fontfamily{ptm}\selectfont \textit{age}} distribution of clusters.  However, the bottom panel demonstrates how off the $\chi^{2}$ minimization method is in determining age.  Here, the five very red clusters are pushed to be older than $log(age/yr)=9.00$. An additional cluster is presented in the Figure \ref{fig:paper4-13new}, $BDSL\phn 1938$, which is very red only in $(V-R)_{0}$. Its $\chi^{2}$ minimization age is younger than the CMD or the MASSCLEAN{\fontfamily{ptm}\selectfont \textit{age}} age, bellow $log(age/yr)=7.0$. This example shows how the $\chi^{2}$ minimization method might be producing an over-density of younger and older clusters, with an under-density of clusters in the $log(age/yr)=[7.0,7.5]$ range.

\begin{deluxetable}{lrrrrrrrrrrrr}
\tablecolumns{13}
\tablewidth{0pc}
\tabletypesize{\scriptsize}
\tablecaption{Age Determination for 6 Red Clusters \label{table3}}
\tablehead{
\colhead{}    &  \multicolumn{4}{c}{{\scriptsize Integrated Photometry (Hunter et al. 2003)}} & \colhead{} & \colhead{} & \colhead{} & \colhead{} & \colhead{}  & \multicolumn{2}{c}{{\scriptsize MASSCLEAN}} \\
\cline{2-5} \cline{11-12}\\
\colhead{{\scriptsize Name}} & \colhead{{\scriptsize$M_{V}$}} & \colhead{{\scriptsize$(U-B)_{0}$}} & \colhead{{\scriptsize$(B-V)_{0}$}} & \colhead{{\scriptsize$(V-R)_{0}$}}& \colhead{{\tiny }} & \colhead{{\scriptsize Age\tablenotemark{a}}}  & \colhead{{\tiny }} & \colhead{{\scriptsize Age\tablenotemark{b}}}  & \colhead{{\tiny }}  & \colhead{{\scriptsize Age}}  & \colhead{{\scriptsize Mass}} \\
\colhead{{\scriptsize }} & \colhead{{\scriptsize$(mag)$}} & \colhead{{\scriptsize$(mag)$}} & \colhead{{\scriptsize$(mag)$}} & \colhead{{\scriptsize$(mag)$}} & \colhead{{\tiny }} & \colhead{{\scriptsize $(log)$}} & \colhead{{\tiny }} & \colhead{{\scriptsize $(log)$}}  & \colhead{{\tiny }}  &  \colhead{{\scriptsize $(log)$}} & \colhead{{\scriptsize $(M_{\Sun})$}} \\
\colhead{{\tiny$1$}} & \colhead{{\tiny$2$}} & \colhead{{\tiny$3$}} & \colhead{{\tiny$4$}} & \colhead{{\tiny$5$}} & \colhead{{\tiny }} & \colhead{{\tiny$6$}} & \colhead{{\tiny }}  & \colhead{{\tiny$7$}}  & \colhead{{\tiny }} & \colhead{{\tiny$8$}} & \colhead{{\tiny$9$}}   }
\startdata

{\tiny BSDL1943} & {\tiny$-6.824\pm 0.006$  \phn} & {\tiny$-0.268\pm 0.007  $  \phn} & {\tiny$1.052\pm 0.009  $  \phn} & {\tiny$0.920\pm 0.008  $  \phn} & {\tiny} & {\tiny$7.20\pm 0.50$  \phn} & {\tiny} & {\tiny$9.16$  \phn} & {\tiny} & {\tiny$7.21^{+0.02}_{-0.02}   $  \phn} & {\tiny$750^{+250 \phn \phn }_{-150} $  \phn} \\
{\tiny KMHK971} & {\tiny$-6.503\pm 0.006$  \phn} & {\tiny$-0.256\pm 0.004  $  \phn} & {\tiny$0.939\pm 0.007  $  \phn} & {\tiny$0.896\pm 0.008  $  \phn} & {\tiny} & {\tiny$7.40\pm 0.50$  \phn} & {\tiny} & {\tiny$9.16$  \phn} & {\tiny} & {\tiny$7.30^{+0.02}_{-0.01}   $  \phn} & {\tiny$850^{+150 \phn \phn }_{-150} $  \phn} \\
{\tiny BRHT62b} & {\tiny$-5.810\pm 0.009$  \phn} & {\tiny$-0.358\pm 0.007  $  \phn} & {\tiny$0.724\pm 0.011  $  \phn} & {\tiny$0.732\pm 0.014  $  \phn} & {\tiny} & {\tiny$7.50\pm 0.30$  \phn} & {\tiny} & {\tiny$9.25$  \phn} & {\tiny} & {\tiny$7.54^{+0.02}_{-0.01}   $  \phn} & {\tiny$950^{+50 \phn \phn \phn}_{-150} $  \phn} \\
{\tiny BRHT60b} & {\tiny$-5.700\pm 0.010$  \phn} & {\tiny$-0.222\pm 0.009  $  \phn} & {\tiny$0.745\pm 0.013  $  \phn} & {\tiny$0.638\pm 0.016  $  \phn} & {\tiny} & {\tiny$7.60\pm 0.40$  \phn} & {\tiny} & {\tiny$9.25$  \phn} & {\tiny} & {\tiny$7.75^{+0.07}_{-0.11}   $  \phn} & {\tiny$1200^{+300 \phn \phn }_{-300} $  \phn} \\
{\tiny BSDL66} & {\tiny$-5.906\pm 0.010$  \phn} & {\tiny$-0.184\pm 0.007  $  \phn} & {\tiny$0.751\pm 0.012  $  \phn} & {\tiny$0.624\pm 0.016  $  \phn} & {\tiny} & {\tiny$7.60\pm 0.30$  \phn} & {\tiny} & {\tiny$9.25$  \phn} & {\tiny} & {\tiny$7.64^{+0.07}_{-0.06}   $  \phn} & {\tiny$1100^{+200 \phn \phn }_{-100} $  \phn} \\
{\tiny BSDL1938} & {\tiny$-6.728\pm 0.005$  \phn} & {\tiny$-0.632\pm 0.004  $  \phn} & {\tiny$0.489\pm 0.006  $  \phn} & {\tiny$0.722\pm 0.008  $  \phn} & {\tiny} & {\tiny$7.60\pm 0.30$  \phn} & {\tiny} & {\tiny$6.82$  \phn} & {\tiny} & {\tiny$7.34^{+0.02}_{-0.03}   $  \phn} & {\tiny$2000^{+100 \phn \phn }_{-400} $  \phn} \\

\enddata
\tablenotetext{a}{{\scriptsize CMD age from \citeauthor*{glatt} \citeyear{glatt}.}}
\tablenotetext{b}{{\scriptsize Age determined by traditional $\chi^{2}$ minimization based on the {\it infinite mass limit}.}}
\end{deluxetable}


\section{Fading Limit Revisited}\label{fading}

Stellar evolution causes clusters to fade in luminosity with age, as displayed in the Figures \ref{fig:paper4-02new} and \ref{fig:paper4-05new}. The number of observable clusters for a given limiting magnitude (detection limit) will decrease with decreasing cluster mass and increasing age.  There exists an intrinsic {\it fading limit}, in the cluster mass-age space, that manifests in the detection limit in one's imaging surveys.  When stellar clusters are described by SSP models computed in the {\it infinite mass limit}, the fading limit is simply a line in a two-dimensional plot of cluster age and mass.  This line delineates the age limit beyond which a cluster of a certain mass is expected to be too faint to be detected (e.g. \citeauthor*{lamers2003} \citeyear{lamers2003}, \citeauthor*{chandar2010a} \citeyear{chandar2010a}).  However, this ignores the effect stochastic fluctuations have on the observed magnitude of real clusters.  A more realistic fading limit is not simply a line, but a probabilistic {\it region} in a cluster mass - age diagram.  

We have explored this concept using our MASSCLEAN{\fontfamily{ptm}\selectfont \textit{colors}} database.  We assume a limiting magnitude of $M_{V}=-4$ mag and derive a fading limit from {\it 100 million} Monte Carlo simulations.  This fading limit {\it region} is given in gray in Figure \ref{fig:paper4-20} and corresponds to clusters with $M_{V}=-4.000 \pm 0.005$ mag (this magnitude range is much smaller than the photometric errors in $M_{V}$ for our LMC cluster sample).  The traditional fading limit, based on the same limiting magnitude but computed using the {\it infinite mass limit}, is presented as the bold black line.  We have plotted all 920 LMC clusters in Figure \ref{fig:paper4-20}.  Those clusters brighter than the limiting magnitude of  $M_{V}=-4$ mag are represented as red dots, while the clusters fainter than this limiting magnitude are shown as blue dots. The size of the dots is scaled with the $M_{V}$ magnitude.  If our MASSCLEAN simulations are reasonable, no red dots should appear below the gray region, and no blue dots should appear above the gray region.  Within the gray region, red or blue dots will co-exist. This is true in Figure \ref{fig:paper4-20}, except for a few red dots lying below the gray region between $log(age/yr) = 7.50 - 7.75$, and a few very old, low mass clusters appearing as red dots below the gray region above $log(age/yr)= 9.20$.  It is not clear whether the gray region we've determined or the age and mass of those few clusters (using MASSCLEAN{\fontfamily{ptm}\selectfont \textit{age}}) are the cause (and it may be both). 

Despite this, Figure \ref{fig:paper4-20} does demonstrate our point.  Some (mostly low mass) clusters located in the gray area could be brighter than the $M_{V}=-4$ mag observed limiting magnitude (red dots), while other clusters with a similar age and mass could be fainter than the  observed limiting magnitude (blue dots).   A fraction of clusters located below the traditional fading limit will exist above the observed limiting magnitude and will indeed be detected and a fraction of the clusters located above the traditional fading limit, are in reality below the observed magnitude limit and will not be detected. As expected, Figure \ref{fig:paper4-20} demonstrates that this effect is prominent with low mass clusters (below $10^{3}$ $M_{\Sun}$).   But, it also shows that young cluster are more susceptible to this effect, as the gray fading region becomes very wide for younger clusters, $log(age/yr) < 8.2$.  

\begin{figure}[htp]
\centering
\includegraphics[angle=270,width=8.5cm, bb= 120 110 515 670]{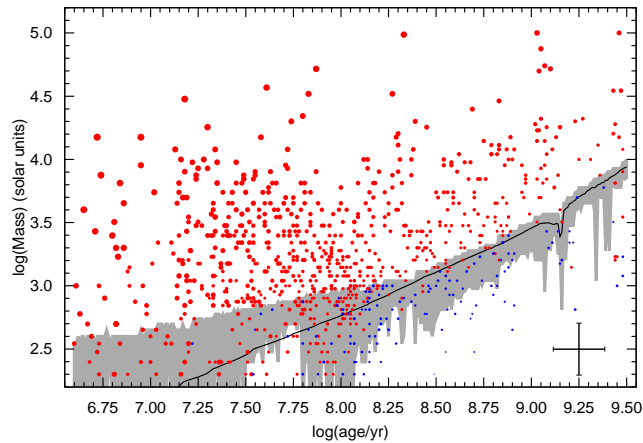}
\caption[]{\footnotesize The LMC clusters brighter than the $M_{V}=-4 \phn mag$ observed magnitude limit are represented as red dots, while the clusters fainter than this limit are shown as blue dots. The size of the dots is scaled with the $M_{V}$ magnitude. The black line represents a traditional fading line. The gray zone represents the range over which clusters of that mass and age may or may not be seen to exceed the $M_V=-4 \phn mag$ observed limiting magnitude.\normalsize}\label{fig:paper4-20}
\end{figure}

The MASSCLEAN{\fontfamily{ptm}\selectfont \textit{age}} results for mass and age are presented as histograms in Figures \ref{fig:paper4-21}-\ref{fig:paper4-22}. The entire distribution of 920 LMC clusters is presented here.  The clusters brighter than the $M_{V}=-4 \phn mag$ magnitude limit, shown as red dots in Figure \ref{fig:paper4-20}, are now displayed in black.  The clusters dimmer than the $M_{V}=-4 \phn mag$ magnitude limit, shown as blue dots in Figure \ref{fig:paper4-20}, are displayed in gray.  

We wish to derive a mathematically description of the distribution of LMC clusters in our sample as a function of age and mass.  These histograms demonstrate that a power-law fit to age and mass should be possible, provided we limit ourselves to the black areas, meaning masses over $10^3$ $M_{\Sun}$ and ages greater than $log(age/yr) > 8.0$.  This means most of the gray region of Figure \ref{fig:paper4-21} and Figure \ref{fig:paper4-22} will not be included in our final distribution.  Our final distribution, which includes all clusters in Figures \ref{fig:paper4-21} and \ref{fig:paper4-22} that are both greater than these two limits, when fit to the generic formula, $d^{2}N/dM dt \propto M^{\alpha}t^{\beta}$, give the result $\alpha = -1.5$ to $-1.6$ and $\beta = -2.1$ to $-2.2$.   Because we have attempted to avoid the gray regions of Figures 17 and 18, we need not apply a fading limit in determining this distribution.  It provides nothing more than a mathematical description of our cluster sample.

\begin{figure}[htp]
\centering
\subfigure[]{\includegraphics[angle=270,width=0.48\textwidth, bb= 65 75 554 770]{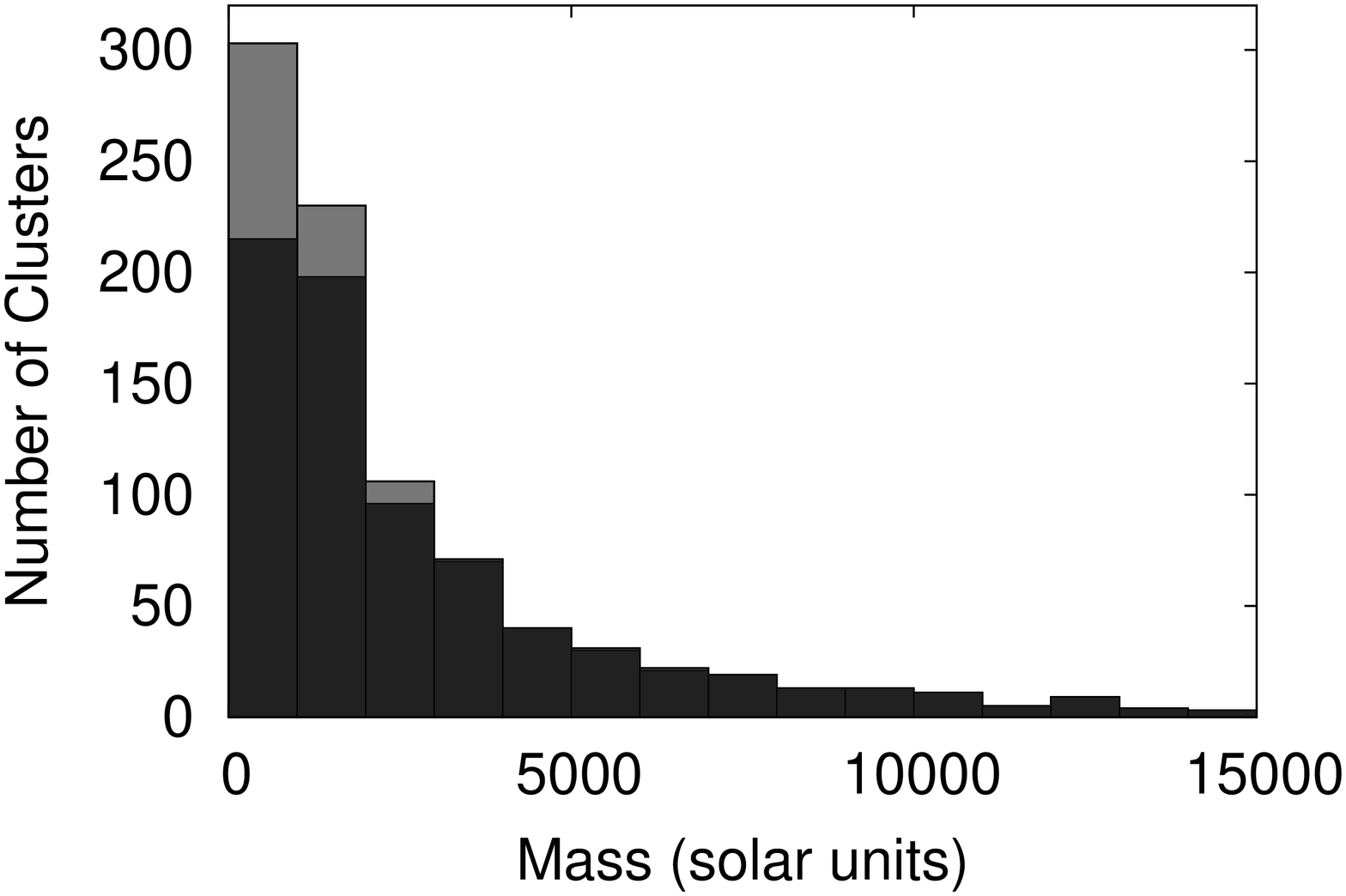}} 
\subfigure[]{\includegraphics[angle=270,width=0.48\textwidth, bb= 65 75 554 770]{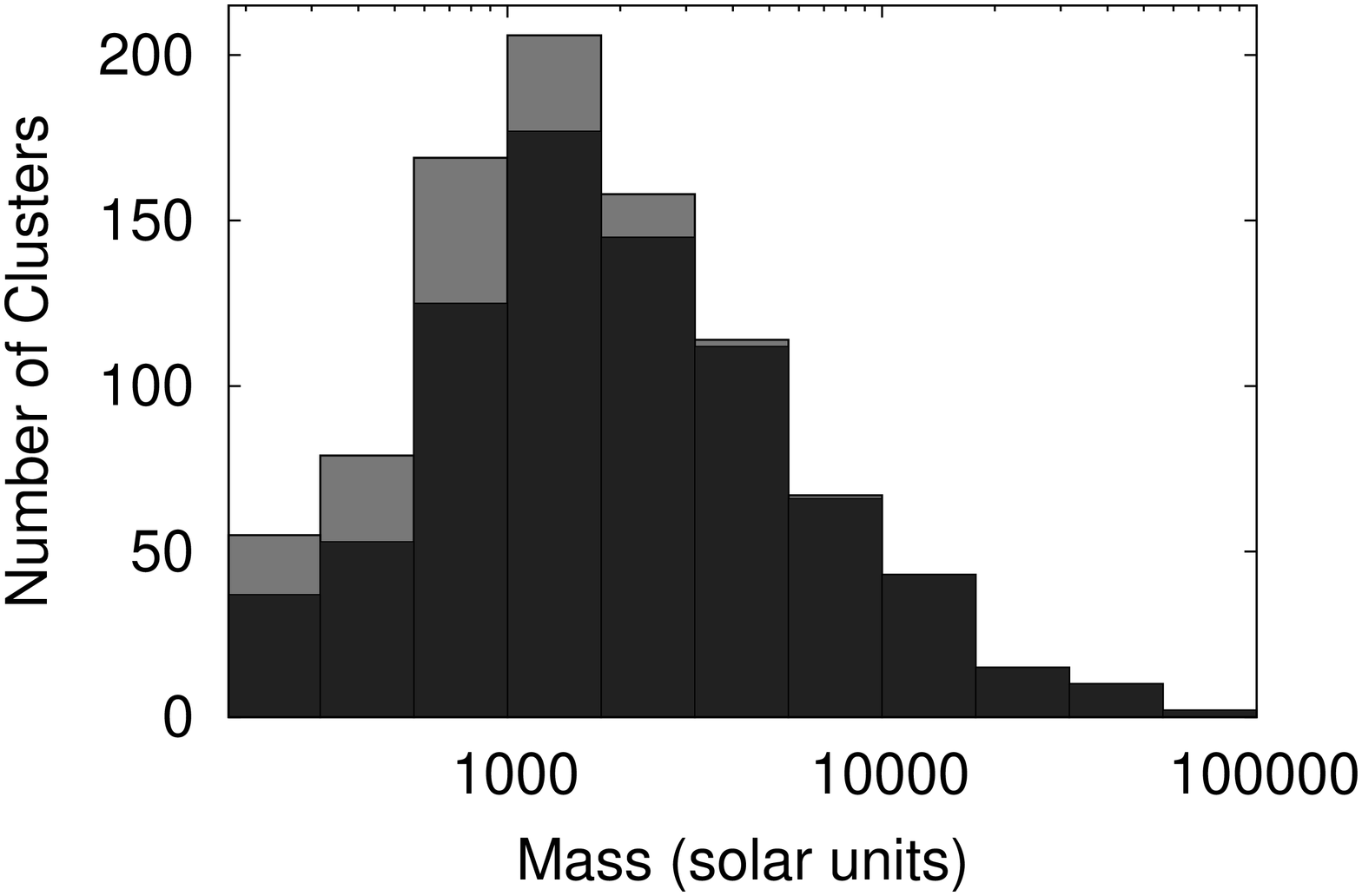}}
\caption[]{\footnotesize Number distribution of 920 LMC clusters based on mass.  Black represents those clusters with intrinsic luminosity brighter than $M_V = -4 \phn mag$; gray clusters lie below this limit. The mean error is given in the lower right corner. \normalsize}\label{fig:paper4-21}
\end{figure}

\begin{figure}[htp]
\centering
\subfigure[]{\includegraphics[angle=270,width=0.48\textwidth, bb= 65 75 554 770]{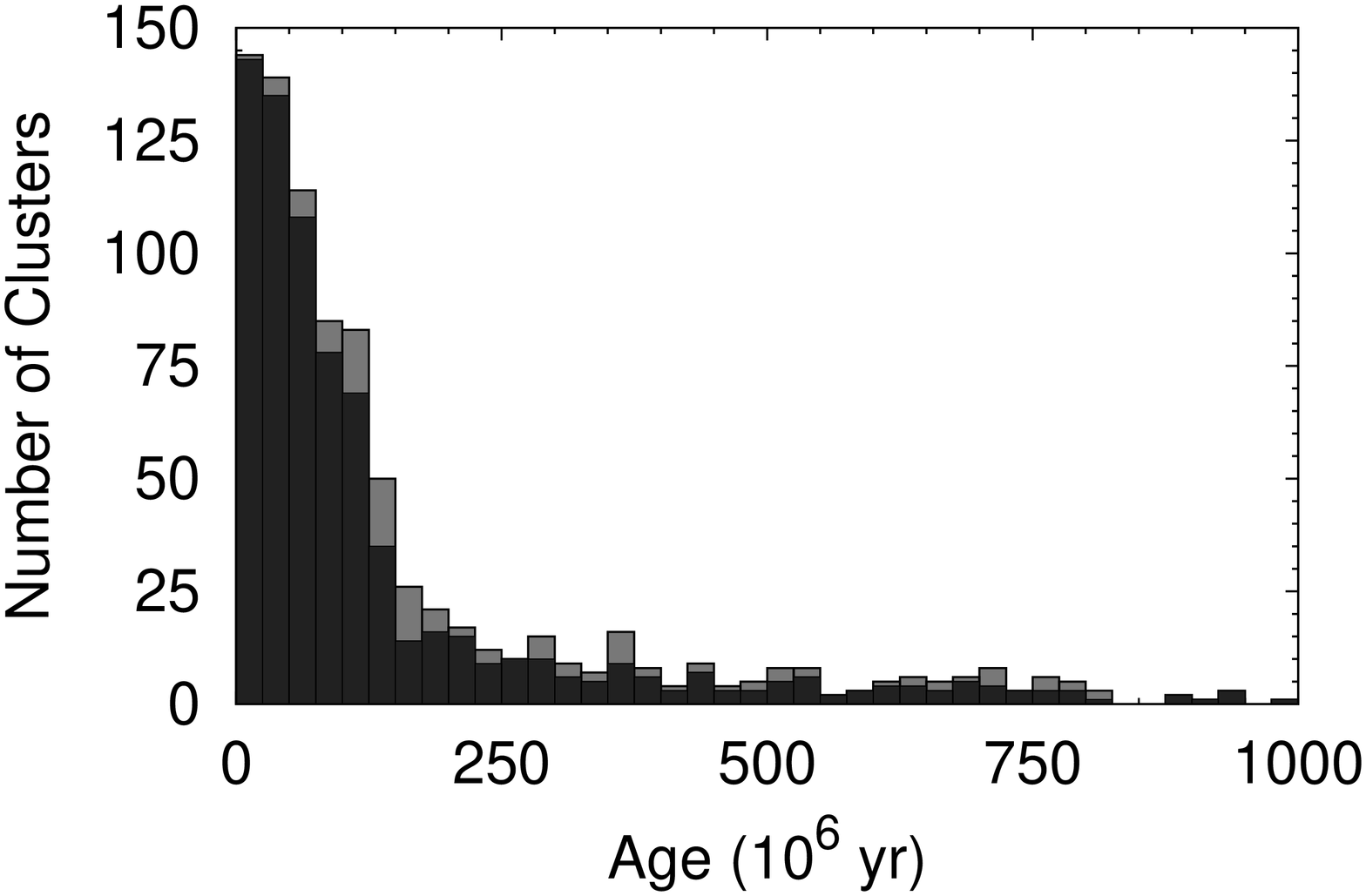}} 
\subfigure[]{\includegraphics[angle=270,width=0.48\textwidth, bb= 65 75 554 770]{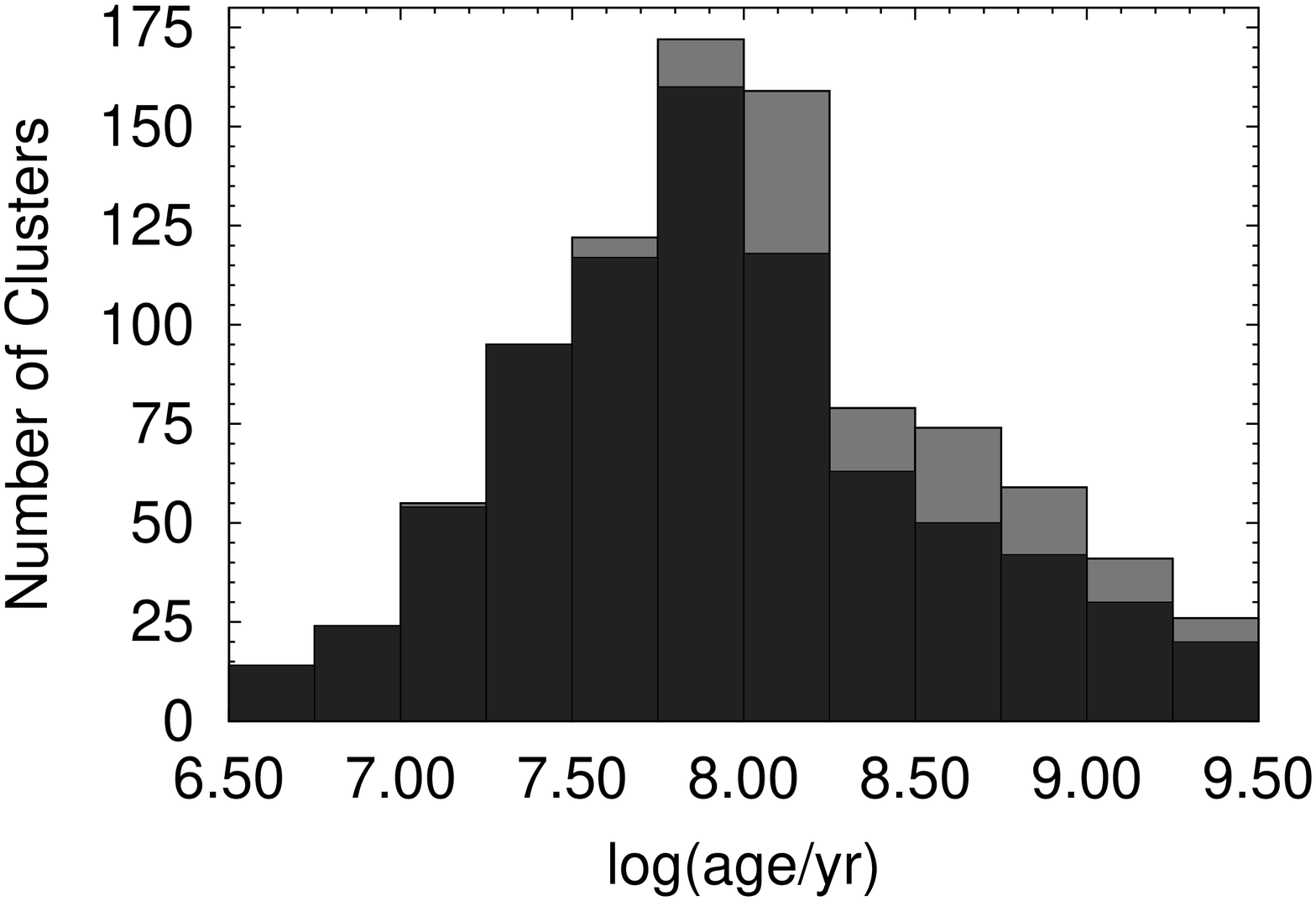}}
\caption[]{\footnotesize Number distribution of  920 LMC clusters based on age.  Black represents those clusters with intrinsic luminosity $M_V = -4$; gray clusters lie below this limit. \normalsize}\label{fig:paper4-22}
\end{figure}

%
%

In order to derive the disruption function for stellar clusters, one needs first a reliable cluster distribution. We have that now for the LMC.  But converting from a cluster distribution to a disruption function is not straight forward.  It requires an understanding of the fading limit in the sample.  We've shown that the traditional method, that does not consider the statistical fluctuations in cluster properties (most relevant here, absolute magnitude) will not provide an accurate representation of the clusters that should be counted. 

At this time, we choose only to display the mass and age distribution for the LMC clusters and not to make any assumptions about the disruption timescale. Instead, we will wait and address the disruption problem in a consistent way, using our mass-dependent SSP models which properly accounts for the stochastic fluctuations as a function of age and mass (as we did in this work), properly applied to the fading limit, and ultimately the disruption.  Mass-dependent SSP models that include stochastic fluctuations for the fading limit and disruption timescale will require an even larger number of Monte Carlo simulations of a population of clusters than we have completed here (e.g. \citeauthor*{parmentier} \citeyear{parmentier}; \citeauthor*{chandar2010b} \citeyear{chandar2010b}; \citeauthor*{oey} \citeyear{oey}).  This will instead be the subject of a future work (Popescu et al. 2012).

\section{Conclusion}

Obtaining accurate, consistent ages and masses for a large set of stellar clusters is a first, critical step to deriving fundamental characteristics about stellar clusters, such as the cluster mass function, the cluster formation rate, and cluster disruption timescales.   To this end, we have completed an analysis to derive the age and mass of 920 clusters, using previously published photometry and our stellar cluster analysis package, MASSCLEAN.  

Our set of results is quite different from previous work. Our age and mass results are not based on integrated colors computed in the {\it infinite mass limit}, as described in section Section \ref{traditional}. For the first time, ages and masses were computed using mass-dependent integrated colors, which include stochastic fluctuations for typical stellar clusters. Since the majority of the clusters have a mass smaller than $10^{4}$ $M_{\Sun}$, our 920 LMC cluster sample is indeed dominated by these stochastic fluctuations. Comparing our MASSCLEAN{\fontfamily{ptm}\selectfont \textit{age}} results for a subset of clusters with recently obtained CMD ages shows the MASSCLEAN{\it ages} to be fairly robust.  

At a sufficient enough distance, the effects of stochastic fluctuations can be assumed to be small.  This is because the absolute magnitude of a detected cluster becomes increasingly brighter at further distances. At a sufficient distance, all detected clusters would be massive enough to be sufficiently populated even at the high mass end, making corrections for stochastic fluctuations no longer necessary. We believe, based on our 100 million Monte Carlo simulations, that stochastic fluctuations are no longer a significant source of observed variation in $U,B,V$ colors once the cluster has an absolute magnitude $M_V = -10.0$, corresponding to a $M = 50,000 M_{\Sun}$ cluster at a distance of $1-100$ $Mpc$. 

Finally, we consider the effect stochastic fluctuations have on the fading limit for stellar cluster surveys.  We show that for a selected survey depth, the corresponding fading limit is not a single line with mass and age.  Due to stochastic variations leading to variations in the apparent magnitude of a cluster, a more complex fading limit region, instead of a line, exists where clusters of similar age and mass might or might not be detected.  We give preliminary results demonstrating this.  However, to fully assess the statistical nature of this effect will require even more Monte Carlo simulations than what we have completed to date.  An analysis of the fading limit, applied to our LMC cluster sample, with the final goal of deriving the disruption timescales as a function of age and mass, will be considered in a follow up study using several hundred million stellar cluster simulations.

\acknowledgements
We are grateful to suggestions made on an early draft of this work by Deidre Hunter, Jay Gallagher, and Sally Oey.  Their ideas lead to significant improvements in the presentation. We thank the referee for useful comments and suggestions. 
This material is based upon work supported by the National Science Foundation under grant AST-0607497 and AST-1009550, to the University of Cincinnati. This work also was funded in part by the National Science Foundation through grant AST-0707426 to BGE.

\begin{deluxetable}{lrrrrrrrrrrr}
\rotate
\tablecolumns{12}
\tablewidth{0pt}
\tabletypesize{\scriptsize}
\tablecaption{Age and Mass Determination for 288 LMC Clusters with CMD Age \label{table1}}
\tablehead{
\colhead{}    &  \multicolumn{5}{c}{{\scriptsize Integrated Photometry (Hunter et al. 2003)}} & \colhead{{\tiny }}     &  \multicolumn{2}{c}{{\scriptsize CMD (Glatt et al. 2010) }}    & \colhead{{\tiny }} &   \multicolumn{2}{c}{{\scriptsize MASSCLEAN}} \\
\cline{2-6} \cline{8-9} \cline{11-12}\\
\colhead{{\scriptsize Name(s)}} & \colhead{{\scriptsize$M_{V}$}} & \colhead{{\scriptsize$(U-B)_{0}$}} & \colhead{{\scriptsize$(B-V)_{0}$}} & \colhead{{\scriptsize$(V-R)_{0}$}} & \colhead{{\scriptsize Age}}  & \colhead{{\tiny }}  & \colhead{{\scriptsize Age}}  & \colhead{{\scriptsize E(B-V)}} & \colhead{{\tiny }}  & \colhead{{\scriptsize Age}}  & \colhead{{\scriptsize Mass}} \\
\colhead{{\scriptsize }} & \colhead{{\scriptsize$(mag)$}} & \colhead{{\scriptsize$(mag)$}} & \colhead{{\scriptsize$(mag)$}} & \colhead{{\scriptsize$(mag)$}} & \colhead{{\scriptsize $(log)$}}  & \colhead{{\tiny }}  &  \colhead{{\scriptsize $(log)$}} & \colhead{{\scriptsize $(mag)$}}  & \colhead{{\tiny }}  &  \colhead{{\scriptsize $(log)$}} & \colhead{{\scriptsize $(M_{\Sun})$}} \\
\colhead{{\tiny$1$}} & \colhead{{\tiny$2$}} & \colhead{{\tiny$3$}} & \colhead{{\tiny$4$}} & \colhead{{\tiny$5$}} & \colhead{{\tiny$6$}} & \colhead{{\tiny }}  & \colhead{{\tiny$7$}} & \colhead{{\tiny$8$}} & \colhead{{\tiny }}  & \colhead{{\tiny$9$}} & \colhead{{\tiny$10$}} 

}
\startdata

{\tiny BSDL305} & {\tiny$-7.801\pm 0.004$  \phn} & {\tiny$-0.941\pm 0.002  $  \phn} & {\tiny$-0.083\pm 0.004  $  \phn} & {\tiny$-0.015\pm 0.008  $  \phn} & {\tiny$6.61  $  \phn} & {\tiny} & {\tiny$7.00\pm 0.50$  \phn} & {\tiny$0.100$  \phn} & {\tiny} & {\tiny$6.63^{+0.27}_{-0.03}   $  \phn} & {\tiny$600^{+3100 \phn }_{-400} $  \phn} \\
{\tiny NGC1755.SL99} & {\tiny$-9.628\pm 0.002$  \phn} & {\tiny$-0.526\pm 0.001  $  \phn} & {\tiny$-0.141\pm 0.002  $  \phn} & {\tiny$0.000\pm 0.005  $  \phn} & {\tiny$7.21  $  \phn} & {\tiny} & {\tiny$7.40\pm 0.30$  \phn} & {\tiny$0.300$  \phn} & {\tiny} & {\tiny$6.65^{+0.23}_{-0.05}   $  \phn} & {\tiny$4000^{+6000 \phn }_{-2500} $  \phn} \\
{\tiny BSDL1834} & {\tiny$-7.016\pm 0.005$  \phn} & {\tiny$-1.040\pm 0.002  $  \phn} & {\tiny$-0.072\pm 0.005  $  \phn} & {\tiny$0.067\pm 0.010  $  \phn} & {\tiny$6.52  $  \phn} & {\tiny} & {\tiny$7.30\pm 0.40$  \phn} & {\tiny$0.100$  \phn} & {\tiny} & {\tiny$6.69^{+0.24}_{-0.02}   $  \phn} & {\tiny$400^{+1600 \phn }_{-200} $  \phn} \\
{\tiny H88-267} & {\tiny$-9.949\pm 0.001$  \phn} & {\tiny$-1.025\pm 0.001  $  \phn} & {\tiny$0.308\pm 0.001  $  \phn} & {\tiny$0.246\pm 0.002  $  \phn} & {\tiny$6.92  $  \phn} & {\tiny} & {\tiny$7.20\pm 0.40$  \phn} & {\tiny$0.200$  \phn} & {\tiny} & {\tiny$6.72^{+0.27}_{-0.03}   $  \phn} & {\tiny$15000^{+10000 }_{-8000} $  \phn} \\
{\tiny SL557.BRHT15a} & {\tiny$-6.606\pm 0.008$  \phn} & {\tiny$-0.897\pm 0.004  $  \phn} & {\tiny$-0.251\pm 0.009  $  \phn} & {\tiny$-0.058\pm 0.016  $  \phn} & {\tiny$10.00  $  \phn} & {\tiny} & {\tiny$7.60\pm 0.40$  \phn} & {\tiny$0.150$  \phn} & {\tiny} & {\tiny$6.73^{+0.17}_{-0.12}   $  \phn} & {\tiny$1000^{+300 \phn \phn }_{-800} $  \phn} \\
\enddata
\end{deluxetable}

\begin{deluxetable}{lrrrrrrrr}
\tablecolumns{9}
\tablewidth{0pc}
\tabletypesize{\scriptsize}
\tablecaption{Age and Mass Determination for 632 LMC Clusters\label{table2}}
\tablehead{
\colhead{}    &  \multicolumn{5}{c}{{\scriptsize Integrated Photometry (Hunter et al. 2003)}} & \colhead{}  & \multicolumn{2}{c}{{\scriptsize MASSCLEAN}} \\
\cline{2-6} \cline{8-9}\\
\colhead{{\scriptsize Name(s)}} & \colhead{{\scriptsize$M_{V}$}} & \colhead{{\scriptsize$(U-B)_{0}$}} & \colhead{{\scriptsize$(B-V)_{0}$}} & \colhead{{\scriptsize$(V-R)_{0}$}} & \colhead{{\scriptsize Age}}  & \colhead{{\tiny }}  & \colhead{{\scriptsize Age}}  & \colhead{{\scriptsize Mass}} \\
\colhead{{\scriptsize }} & \colhead{{\scriptsize$(mag)$}} & \colhead{{\scriptsize$(mag)$}} & \colhead{{\scriptsize$(mag)$}} & \colhead{{\scriptsize$(mag)$}} & \colhead{{\scriptsize $(log)$}}  & \colhead{{\tiny }}  &  \colhead{{\scriptsize $(log)$}} & \colhead{{\scriptsize $(M_{\Sun})$}} \\
\colhead{{\tiny$1$}} & \colhead{{\tiny$2$}} & \colhead{{\tiny$3$}} & \colhead{{\tiny$4$}} & \colhead{{\tiny$5$}} & \colhead{{\tiny$6$}} & \colhead{{\tiny }}  & \colhead{{\tiny$7$}} & \colhead{{\tiny$8$}}   }
\startdata

{\tiny KMHK237} & {\tiny$-6.638\pm 0.006$  \phn} & {\tiny$-0.979\pm 0.002  $  \phn} & {\tiny$-0.259\pm 0.006  $  \phn} & {\tiny$-0.143\pm 0.014  $  \phn} & {\tiny$10.00  $  \phn} & {\tiny} & {\tiny$6.60^{+0.15}_{-0.00}   $  \phn} & {\tiny$350^{+750 \phn \phn }_{-100} $  \phn} \\
{\tiny BSDL635} & {\tiny$-7.666\pm 0.004$  \phn} & {\tiny$-1.164\pm 0.002  $  \phn} & {\tiny$-0.020\pm 0.004  $  \phn} & {\tiny$-0.069\pm 0.008  $  \phn} & {\tiny$6.76  $  \phn} & {\tiny} & {\tiny$6.61^{+0.05}_{-0.01}   $  \phn} & {\tiny$1000^{+1000 \phn }_{-500} $  \phn} \\
{\tiny H88-266} & {\tiny$-7.077\pm 0.009$  \phn} & {\tiny$-1.055\pm 0.004  $  \phn} & {\tiny$-0.206\pm 0.009  $  \phn} & {\tiny$-0.064\pm 0.017  $  \phn} & {\tiny$6.53  $  \phn} & {\tiny} & {\tiny$6.67^{+0.09}_{-0.01}   $  \phn} & {\tiny$500^{+700 \phn \phn }_{-250} $  \phn} \\
{\tiny BSDL137} & {\tiny$-5.099\pm 0.014$  \phn} & {\tiny$-1.223\pm 0.004  $  \phn} & {\tiny$-0.270\pm 0.014  $  \phn} & {\tiny$-0.080\pm 0.033  $  \phn} & {\tiny$6.36  $  \phn} & {\tiny} & {\tiny$6.68^{+0.06}_{-0.07}   $  \phn} & {\tiny$250^{+50 \phn \phn \phn}_{-50} $  \phn} \\
{\tiny BSDL1109} & {\tiny$-9.175\pm 0.002$  \phn} & {\tiny$-0.330\pm 0.001  $  \phn} & {\tiny$0.355\pm 0.002  $  \phn} & {\tiny$0.067\pm 0.004  $  \phn} & {\tiny$6.92  $  \phn} & {\tiny} & {\tiny$6.71^{+0.29}_{-0.01}   $  \phn} & {\tiny$2700^{+5300 \phn }_{-1300} $  \phn} \\
{\tiny BSDL295} & {\tiny$-6.303\pm 0.008$  \phn} & {\tiny$-0.927\pm 0.003  $  \phn} & {\tiny$-0.268\pm 0.008  $  \phn} & {\tiny$-0.133\pm 0.018  $  \phn} & {\tiny$10.00  $  \phn} & {\tiny} & {\tiny$6.72^{+0.03}_{-0.01}   $  \phn} & {\tiny$300^{+200 \phn \phn }_{-100} $  \phn} \\
{\tiny BSDL2208} & {\tiny$-5.618\pm 0.012$  \phn} & {\tiny$-0.894\pm 0.005  $  \phn} & {\tiny$-0.296\pm 0.012  $  \phn} & {\tiny$-0.086\pm 0.026  $  \phn} & {\tiny$6.62  $  \phn} & {\tiny} & {\tiny$6.73^{+0.01}_{-0.01}   $  \phn} & {\tiny$200^{+200 \phn \phn }_{-0} $  \phn} \\
\enddata
\end{deluxetable}


\begin{thebibliography}{}

\small 
\bibitem[Asa'd \& Hanson(2012)Asa'd \& Hanson]{asa'd} Asa'd, R.S., Hanson, M.M. 2012, MNRAS, 419, 2116
\bibitem[Bastian et al.(2005)Bastian et al.]{bast05} Bastian, N., Gieles, M., Lamers, H. J. G. L. M., Scheepmaker, R. A., de Grijs, R. 2005, A\&A, 431, 905
\bibitem[Bastian et al.(2011)Bastian et al.]{bast11} Bastian, N., Adamo, A., Gieles, M., Lamers, H. J. G. L. M., Larsen, S. S., Silva-Villa, E., Smith, L. J., Kotulla, R., Konstantopoulos, I. S., Trancho, G., Zackrisson, E. 2011, MNRAS, 417, L6
\bibitem[Bastian et al.(2012)Bastian et al.]{bast12} Bastian, N., Adamo, A., Gieles, M., Silva-Villa, E., Lamers, H. J. G. L. M., Larsen, S. S., Smith, L. J., Konstantopoulos, I. S., Zackrisson, E. 2012, MNRAS, 419, 2606 
\bibitem[Bica et al. (2008)Bica et al.]{bica2008} Bica, E., Bonatto, C., Dutra, C. M., Santos, J. F. C. 2008, MNRAS, 389, 678
\bibitem[Bik et al. (2003)Bik et al.]{bik2003} Bik, A., Lamers, H. J. G. L. M., Bastian, N., Panagia, N., Romaniello, M. 2003, A\&A, 397, 473
\bibitem[Boutloukos \& Lamers (2003)Boutloukos \& Lamers]{lamers2003} Boutloukos, S.G. \& Lamers, H.J.G.L.M. 2003, MNRAS, 338, 717
\bibitem[Cardelli et al. (1989)Cardelli, Clayton and Mathis]{ccm} Cardelli, J. A., Clayton, G. C., Mathis, J. S. 1989, \apj, 345, 245
\bibitem[Chandar et al. (2010a)Chandar et al.]{chandar2010a} Chandar, R., Fall, S.M, Whitmore, B.C. 2010a, ApJ, 711, 1263
\bibitem[Chandar et al. (2010b)Chandar et al.]{chandar2010b} Chandar, R., Whitmore, B.C., Fall, S.M. 2010b, ApJ, 713, 1343
\bibitem[da Silva et al.(2012)da Silva et al.]{slug} da Silva, R. L., Fumagalli, M., Krumholz, M. 2012, ApJ, 745, 145
\bibitem[de Grijs et al.(2003)de Grijs et al.]{deGrijs2003} de Grijs, R., Fritze-v. Alvensleben, U., Anders, P., Gallagher, J. S., Bastian, N., Taylor, V. A., Windhorst, R. A. 2003, MNRAS, 342, 259
\bibitem[de Grijs \& Anders (2006)de Grijs \& Anders]{deGrijs2006} de Grijs, R, Anders, P. 2006, MNRAS, 366, 295
\bibitem[Dolphin \& Kennicutt(2002)Dolphin \& Kennicutt]{Dolphin2002} Dolphin, A.E., Kennicutt, R.C. 2002, AJ, 124, 158
\bibitem[Fall et al. (2005)Fall et al.]{fall2005} Fall, S.M., Chandar, R., Whitmore, B.C. 2005 \apjl, 631, L133
\bibitem[Fall et al. (2009)Fall et al.]{fall2009} Fall, S.M., Chandar, R., Whitmore, B.C. 2009 \apj, 704, 453
\bibitem[Fouesneau \& Lan{\c c}on (2010)Fouesneau \& Lan{\c c}on]{fouesneau2} Fouesneau, M. \& Lan{\c c}on, A. 2010, A\&A, 521, 22
\bibitem[Gieles et al. (2005)Gieles et al.]{Gieles2005} Gieles, M., Bastian, N., Lamers, H. J. G. L. M., Mout, J. N. 2005, A\&A, 441, 949
\bibitem[Girardi et al. (2010)Girardi et al.]{padova2010} Girardi, L., Bressan, A., Bertelli, G., Chiosi, C. 2010 A\&AS, 141, 371
\bibitem[Glatt et al. (2010)Glatt et al.]{glatt}Glatt, K., Grebel, E.K., Koch, A. 2010 A\&A, 517, 50
\bibitem[Hancock et al. (2008)Hancock et al.]{hancock} Hancock, M., Smith, B.J., Giroux, M.L., Struck, C. 2008, MNRAS 389, 1470
\bibitem[Harris \& Zaritsky (2009)Harris \& Zaritsky]{harris} Harris, J. \& Zaritsky, D. 2009, AJ, 138, 1243
\bibitem[Hunter et al. (2003)Hunter et al.]{hunter2003} Hunter, D.A., Elmegreen, B.G., Dupuy, T.J., Mortonson, M. 2003, \aj, 126, 1836
\bibitem[Kaleida \& Scowen (2010)Kaleida \& Scowen]{Kaleida2010} Kaleida, C. \& Scowen, P.A., 2010, AJ, 140, 379
\bibitem[Kroupa (2002)Kroupa]{Kroupa2002} Kroupa, P. 2002, {\it Sci}, 295, 82
\bibitem[Lada \& Lada (2003)Lada \& Lada]{lada} Lada, C.J, \& Lada, E.A. 2003, ARA\&A, 41, 57
\bibitem[Lamers et al.(2005)Lamers et al.]{lamers2005}Lamers, H. J. G. L. M., Gieles, M., Bastian, N., Baumgardt, H., Kharchenko, N. V., Portegies Zwart, S. 2005, A\&A, 441, 117
\bibitem[Lamers et al.(2010)Lamers et al.]{lamers2010}Lamers, H. J. G. L. M., Baumgardt, H., Gieles, M. 2010, MNRAS, 409, 305
\bibitem[Lan{\c c}on \& Mouhcine (2000)Lan{\c c}on \& Mouhcine]{lancon2000} Lan{\c c}on, A. \& Mouhcine, M. 2000, ASPC, 211, 34
\bibitem[Lan{\c c}on \& Mouhcine (2002)Lan{\c c}on \& Mouhcine]{lancon2002} Lan{\c c}on, A. \& Mouhcine, M. 2002, A\&A, 393, 167
\bibitem[Lan{\c c}on \& Fouesneau (2010)Lan{\c c}on \& Fouesneau]{lancon2010} Lan{\c c}on, A. \& Fouesneau, M. 2010, ASPC, 425, 55
\bibitem[Lan{\c c}on (2011)Lan{\c c}on]{lancon2011} Lan{\c c}on, A. 2011, ASPC, 445, 379
\bibitem[Larsen (2010)Larsen]{larsen2010} Larsen, S. S. 2010, RSPTA, 368, 867
\bibitem[Lejeune \& Schaerer(2001)Lejeune \& Schaerer]{geneva1} Lejeune, T. \& Schaerer, D. 2001, \aap, 366, 538
\bibitem[Marigo et al.(2008)Marigo et al.]{padova2008} Marigo, P., Girardi, L., Bressan, A., Groenewegen, M. A. T., Silva, L., Granato, G. L. 2008, A\&A, 482, 833
\bibitem[Massey (2002)Massey]{massey2002} Massey, P. 2002, \apjs, 141, 81
\bibitem[Oey (2011)Oey]{oey} Oey, M.S. 2011, \apj, 739, L46
\bibitem[Parmentier \& de Grijs(2008)Parmentier \& de Grijs]{parmentier} Parmentier, G. \& de Grijs, R. 2008, MNRAS, 383, 1103
\bibitem[Popescu \& Hanson (2009)Popescu \& Hanson]{paper1} Popescu, B. \& Hanson, M.M. 2009, \aj, 138, 1724
\bibitem[Popescu \& Hanson (2010a)Popescu \& Hanson]{paper2} Popescu, B. \& Hanson, M.M. 2010a,\apj, 713, L21 
\bibitem[Popescu \& Hanson (2010b)Popescu \& Hanson]{paper3} Popescu, B. \& Hanson, M.M. 2010b, \apj, 724, 296
\bibitem[Pessev et al. (2008)Pessev et al.]{pessev2008} Pessev, P. M., Goudfrooij, P., Puzia, T. H., Chandar, R. 2008, MNRAS, 385, 1535
\bibitem[Rafelski \& Zaritsky (2005)Rafelski \& Zaritsky]{rz05} Rafelski, M. \& Zaritsky, D. 2005, \aj, 129, 2701 
\bibitem[Santos et al.(2006)Santos et al.]{santos} Santos, J.F.C, Clari\'a, J.J., Ahumada, A.V., Bica, E., Piatti, A.E., Parisi, M.C. 2006, A\&A, 448, 1023
\bibitem[Searle, Sargent \& Bagnuolo (1973)Searle, Sargent \& Bagnuolo]{searle1973} Searle, L., Sargent, W.L.W., Bagnuolo, W.G. 1973, \apj, 179, 427
\bibitem[Silva-Villa \& Larsen (2011)Silva-Villa \& Larsen]{esteban} Silva-Villa, E. \& Larsen, S.S. 2011, A\&A, 529, 25
\bibitem[Zaritsky et al.(2002)Zaritsky et al.]{z2002} Zaritsky, D., Harris, J., Thompson, I. B., Grebel, E. K., Massey, P. 2002, \aj, 123, 855
\bibitem[Zaritsky et al.(2004)Zaritsky et al.]{z2004} Zaritsky, D., Harris, J., Thompson, I. B., Grebel, E. K. 2004, \aj, 128, 1606
\normalsize
\end{thebibliography}
\end{document}